\documentclass[11pt,a4paper]{article}
\pdfoutput=1
\usepackage{jheppub}

\newcommand{\be}{\begin{equation}}
\newcommand{\ee}{\end{equation}}
\newcommand{\ba}{\begin{eqnarray}}
\newcommand{\ea}{\end{eqnarray}}
\newcommand{\la}{\langle}
\newcommand{\ra}{\rangle}

\newcommand{\kd}{\kappa^{D}}
\newcommand{\rd}{\rho^{D}}
\newcommand{\ku}{\kappa^{U}}
\newcommand{\ru}{\rho^{U}}
\newcommand{\kl}{\kappa^{L}}
\newcommand{\rl}{\rho^{L}}

\newcommand{\sba}{s_{\beta-\alpha}}
\newcommand{\cba}{c_{\beta-\alpha}}

\title{Constraining General Two Higgs Doublet Models 
by the Evolution of Yukawa Couplings}
\author[a] {Johan Bijnens,}
\author[a] {Jie Lu}
\author[a] {and Johan Rathsman}

\affiliation[a]{Department of Astronomy and Theoretical Physics, Lund University,\\
S\"olvegatan 14A, SE 223-62 Lund, Sweden}

\emailAdd{Johan.Bijnens@thep.lu.se}
\emailAdd{Lu.Jie@thep.lu.se}
\emailAdd{Johan.Rathsman@thep.lu.se}

\abstract{
We study how general two Higgs doublet models can be constrained by considering their properties under renormalization group evolution of the Yukawa couplings. We take into account both the appearance of a Landau pole as well as off-diagonal Yukawa couplings leading to flavour changing neutral currents in violation with experimental constraints at the electroweak scale. We find that the latter condition can be used to limit the amount of $Z_2$ symmetry breaking allowed in a given model.
}


\arxivnumber{1111.5760}

\begin{document}
\maketitle
\section{Introduction}
\label{sec:intro}

The Standard Model (SM) has been compared to experiments with great success in the past decades and finding the Higgs boson is the only missing piece. However, there are still a few internal problems. The prime example is the so called hierarchy problem: why is the electroweak (EW) scale much smaller than the Plank scale? Thus, the SM cannot be seen as a fundamental theory of particle physics, but only as an effective description which will break down at higher energies, at least at the Planck scale where gravity becomes of the same magnitude as the gauge forces. The mission of the Large Hadron Collider at CERN is therefore not only to look for the SM Higgs boson but also for physics Beyond the Standard Model (BSM).

The general two Higgs Doublet Model (2HDM) was one of the earliest BSM models, proposed by T.D.~Lee \cite{Lee:1973iz} already in 1973 as a model with spontaneous CP-violation. The 2HDM itself cannot give any solution to the problems of the SM, such as the hierarchy problem. On the contrary, it introduces more problems such as tree level flavour-changing-neutral-currents (FCNC) which are absent in the SM. However, a 2HDM is part of many other BSM models, especially supersymmetric ones, which require an even number of Higgs doublets. 
Therefore it is useful and interesting to study the 2HDM itself, since it can be thought of as an effective description of more general models at the TeV scale. One such example is the Minimal SuperSymmetric Model (MSSM) in the case of heavy superpartners such that the Higgs bosons only decays to SM particles.

The problem of tree level FCNC can be evaded by introducing an appropriate $Z_2$ symmetry that ensures that each fermion type only couples to one of the Higgs doublets, which is sufficient in order to 
avoid tree-level FCNC as shown by Glashow and Weinberg~\cite{Glashow:1976nt}. This is precisely what happens in the MSSM  whose Higgs sector at tree-level is a so called type II 2HDM, meaning that one of the Higgs doublets couples only to  down-type fermions and  the other only to up-type ones. By enforcing a $Z_2$-symmetry one also ensures the absence of tree-level FCNC under renormalization group evolution of the model to other energy scales.

Recently another way of avoiding the tree-level FCNC, by having the Yukawa
couplings to the two Higgs doublets proportional to each other, has been proposed~\cite{Pich:2009sp}. This works fine at a given energy scale but if one evolves the model to another scale then the tree-level FCNC are reintroduced because the Yukawa couplings in this model do not respect any $Z_2$ symmetry as shown by Ferreira et al~\cite{Ferreira:2010xe}. There has also been some discussion of the experimental constraints on this model under renormalization group evolution~\cite{arXiv:1005.5310,arXiv:1005.5706,arXiv:1006.0470} and we will revisit these constraints more carefully below. 

More generally, the FCNC at a given energy scale are avoided as long as the Yukawa couplings are diagonal in the appropriate basis. The constraints on these more general models from low-energy flavour observables have also been studied~\cite{Mahmoudi:2009zx}, but not their properties under renormalization group evolution. 
Apart from these schemes, which are set up in order to avoid tree-level FCNC to a larger or lesser extent, one can also envision a top-down approach where one assumes a certain texture for the mass matrices and from this derives the Yukawa coupling matrices. In the present context the prime example is the Cheng-Sher ansatz \cite{Cheng:1987rs} which gives a natural suppression of tree-level FCNC from the hierarchy of quark masses. Some generic properties of these models under renormalization group evolution have been studied~\cite{cvetic:1997} but not taking experimental constraints into account.

In this paper we will study the properties of all these types of models taking into account also experimental constraints on FCNC when evolving them according to the Renormalization Group Equations (RGE) for the Yukawa couplings. 
In this way we can see how stable the various assumptions are under RGE evolution, which in turn gives a measure of have plausible the assumptions are.  A large sensitivity indicates 
that the assumptions behind the model are not stable meaning that they are either fine-tuned or incomplete such that there for example will be additional particles appearing when going to a higher energy. 
From this respect we will study both the appearance of a Landau pole as well as off-diagonal Yukawa couplings leading to FCNC. Strictly speaking, the experimental constraints on the latter are given at the EW scale. Even so, we can still apply them at a higher scale as a means of a determining the fine-tuning of the models as argued above. As an alternative one can also envision to assume a $Z_2$-symmetric starting point at the EW scale, then evolve up to a high scale where the $Z_2$-symmetry is broken, and finally evolve down to the EW scale again where the experimental FCNC constraints are then applied.

The layout of the paper is as follows.
We first give a brief introduction to the general 2HDM in section \ref{sec:Scalar} including the Yukawa sector with emphasis on the FCNC problem as well as some possible solutions and the RGEs for the Yukawa couplings.
Section \ref{sec:constraints} gives the latest constraints on the non-diagonal Yukawa couplings from neutral meson mixing as well as the SM input values we use. Then in section \ref{sec:analysis}
we present our numerical analysis of the running Yukawa couplings.
We investigate the limits both from the absence of a Landau pole as well as from requiring the off-diagonal Yukawa couplings at higher energy scales to be in accordance with the experimental limits at the EW scale. Finally,
in section \ref{sec:conclusion} we present our conclusions.

\section{The general 2HDM}
\label{sec:Scalar}

\subsection{The Scalar Sector}

The two Higgs doublet model was introduced in \cite{Lee:1973iz} and for a more general overview of its properties and the constraints that can be put on it, we refer to the recent review \cite{Branco:2011iw}. Much of the phenomenology of the 2HDM is also closely related to the SM and MSSM for which we refer to the reviews by Djouadi~\cite{hep-ph/0503172,hep-ph/0503173}.

The most general renormalizable scalar potential with two Higgs doublets, 
$\Phi_1$ and $\Phi_2$, 
can be written as
\begin{eqnarray}
\label{scalarpotential}
V_\Phi &=& m^2_{11}\Phi^\dag_1\Phi_1 + m^2_{22}\Phi^\dag_2\Phi_2
 - (m^2_{12}\Phi^\dag_1\Phi_2 + h.c)\nonumber \\
 &&+ {1\over2}\lambda_1(\Phi_1\Phi_1)^2
 + {1\over2}\lambda_2(\Phi_2\Phi_2)^2
 +\lambda_3(\Phi^\dag_1\Phi_1)(\Phi^\dag_2\Phi_2)
 +\lambda_4(\Phi^\dag_1\Phi_2)(\Phi^\dag_2\Phi_1)\nonumber\\
&& +\Bigg\{{1\over2}\lambda_5(\Phi^\dag_1\Phi_2)^2
+\Big[\lambda_6(\Phi^\dag_1\Phi_1)+\lambda_7(\Phi^\dag_2\Phi_2)\Big](\Phi^\dag_1\Phi_2)
+h.c
\Bigg\}\,.
\end{eqnarray}
The coupling constants $m^2_{11}$, $m^2_{22}$ and $\lambda_{1,2,3,4}$ are real,
while $m_{12}$ and $\lambda_{5,6,7}$ can be complex if there
are not any further restrictions. In the following we will however set them to be real such that there is no explicit CP-violation.

The vacuum expectation values (VEVs) of $\Phi_i$ are in general
\begin{eqnarray}
\label{2HDMvev}
\la\Phi_1\ra_0 &=& {1\over\sqrt2}e^{i\theta_1}\left (
 \begin{array}{c}
0\\
 v_1
\end{array}
\right )\,,\nonumber\\
\la\Phi_2\ra_0 &=&  {1\over\sqrt2}e^{i\theta_2}\left (
 \begin{array}{c}
0\\
v_2
\end{array}
\right )\, ,
\end{eqnarray}
and $\tan\beta$ is defined as the ratio of the $v_i$,
$\tan\beta = v_2/v_1$.

The Higgs doublets can be rotated to a basis in which only one of the doublets
has a vacuum expectation value using the angle $\beta$.
This is called the Higgs basis and is related to the general basis as
\begin{eqnarray}
\label{eq:H1}
H_1& = & \cos\beta\, \Phi_1 + \sin\beta\, e^{-i\theta}\Phi_2\,,\nonumber\\
H_2& = & - \sin\beta\,\Phi_1 + \cos\beta\,e^{-i\theta} \Phi_2\,,
\label{eq:H2}
\end{eqnarray}
with $\theta = \theta_2-\theta_1$.
Hence the VEVs for the doublets in the Higgs basis, with $v^2 = v^2_1+v^2_2$, are
\begin{eqnarray}
\label{eq:Higgsvev}
\la H_1\ra_0 &=& {1\over\sqrt2}e^{i\theta_1}\left (
 \begin{array}{c}
0\nonumber\\
 v
\end{array}
\right )\,,\\
\la H_2\ra_0 &=&  \left (
 \begin{array}{c}
0\\
0
\end{array}
\right )\,.
\end{eqnarray}

We have defined both $\Phi_i$ to have weak hypercharge $+1$.
Doublets with weak hypercharge $-1$ can be constructed out of the complex
conjugate fields via
\be
\widetilde \Phi_i = i\sigma_2 \Phi_i^*\,.
\ee

$\Phi_1$ and $\Phi_2$ consist of 8 real fields in total. Three of them
correspond to the Goldstone bosons to be eaten by 
the weak gauge bosons $W^\pm$ and $Z^0$ upon spontaneous breaking
of the gauge group $SU(2)_L \times U(1)_Y$.
One of the standard
conventions to write the doublets without the Goldstone bosons is (setting for clarity $\theta_1=0$)
\begin{eqnarray}
\label{HiggsDoublets1}
\Phi_1(x) &=& \left (
 \begin{array}{c}
-s_\beta H^+\\
{1\over\sqrt2}( c_\beta v -  s_\alpha h + c_\alpha H - is_\beta A)
\end{array}
\right )\nonumber\\
\label{HiggsDoublets2}
\Phi_2(x) &=&  \left (
 \begin{array}{c}
c_\beta H^+\\
{1\over\sqrt2}(s_\beta v+ c_\alpha h + s_\alpha H + ic_\beta A)
\end{array}
\right )\,.
\end{eqnarray}
Here $H^\pm$ is the charged Higgs boson and
the  angle $\alpha$ ($s_\alpha = \sin\alpha, c_\alpha = \cos\alpha$)
is introduced to diagonalize the CP eigenstates in the neutral sector, which
can be divided into two CP even scalars: ($H$, $h$), and a CP odd pseudo-scalar:
$A$.

\subsection{The Yukawa Sector}
\label{sec:Yukawa}

The weak eigenstates of the SM fermions (with massless neutrinos for simplicity) are denoted as
\begin{eqnarray}
\label{FermionDoublet}
Q_L &=& \left ( \begin{array}{c}U_L\\D_L \end{array}\right )
\qquad
L_L = \left ( \begin{array}{c}\nu_L\\E_L \end{array}\right )\,, \nonumber\\
U_R,&D_R,&E_R\,.
\end{eqnarray}
The most general Yukawa interaction can then be written as
\begin{eqnarray}
-\mathcal{L}_Y&=&
\overline{Q}_L\widetilde{\Phi}_1\eta_1^U U_R+\overline{Q}_L\Phi_1\eta_1^D D_R+\overline{L}_L \Phi_1 \eta_1^L E_R\nonumber\\
&&+\overline{Q}_L\widetilde{\Phi}_2\eta_2^U U_R+\overline{Q}_L \Phi_2\eta_2^D D_R+\overline{L}_L \Phi_2 \eta_2^L E_R+\mathrm{h.c.}\,.
\label{eq:genY}
\end{eqnarray}
We leave the generation index implicit here, all entities are matrices or
vectors in the three-dimensional generation space.
The $\eta^F_i$ are the $3\times3$ matrices of Yukawa couplings for $F=U,D,L$.

In order to show more explicitly the physical content in the Yukawa couplings,
we rotate the Yukawa coupling matrices to the Higgs basis by inverting Eq.~(\ref{eq:H2}) and inserting into Eq.~(\ref{eq:genY}).
\begin{eqnarray}
\label{Yukawa1}
-\mathcal{L}_Y&=&
\overline{Q}_L\widetilde{H}_1\kappa_0^U U_R+\overline{Q}_LH_1\kappa_0^D D_R+\overline{L}_L H_1 \kappa_0^L E_R\nonumber\\
&&+\overline{Q}_L\widetilde{H}_2\rho_0^U U_R+\overline{Q}_L H_2\rho_0^D D_R+\overline{L}_L H_2 \rho_0^L E_R+\mathrm{h.c.}\,.
\end{eqnarray}
The relations between the two sets of Yukawa matrices are
\begin{eqnarray}
\label{eq:kappa}
\kappa_0^U  &=&
\cos\beta\, \eta^U_1 + \sin \beta (e^{-i {\theta}} \eta^U_2) \, ,\nonumber\\
\kappa_0^D  &=&
\cos\beta\, \eta^D_1 + \sin\beta (e^{+i {\theta}} \eta^D_2) \,,\nonumber\\
\kappa_0^L &=&
\cos\beta\, \eta^L_1 + \sin\beta( e^{+i {\theta}} \eta^L_2)\, ;
\end{eqnarray}
and
\begin{eqnarray}
\label{eq:rho}
\rho_0^U & = & - \sin\beta\, \eta^U_1 + \cos\beta
(e^{-i{\theta}} \eta^U_2) \ ,
\nonumber\\
\rho_0^D & = & - \sin\beta\, \eta^D_1 + \cos\beta
(e^{+i{\theta}} \eta^D_2)\,, \nonumber\\
\rho_0^L & = & - \sin\beta\, \eta^L_1 + \cos\beta
(e^{+i{\theta}} \eta^L_2)\,. \label{UDs}
\end{eqnarray}
The couplings to $H_1$ produce the masses of the fermions.
We can go over to the fermion mass basis by bi-diagonalizing the matrices
$\kappa^F$ with the unitary matrices $V_L^F,V^F_R$:
\begin{eqnarray}
\kappa^F &=& V^F_L\kappa_0^F V^{F\dag}_R = {\sqrt{2}\over v}\mathcal{M}_{ii}^F\\
\rho^F &=& V^F_L\rho_0^F V^{F\dag}_R
\label{eq:MasseigenKR}
\end{eqnarray}
The  $\kappa^F$ are diagonal, real
and positive and are fully determined from the fermion masses
$\mathcal{M}_{ii}^F$ with $\mathcal{M}^U_{11} = m_u$ etc.
$\rho^F$ is still a general complex matrix whose non-diagonal matrix elements
could cause tree level flavour-changing-neutral-currents. 
The reason is that we cannot in general diagonalize two different matrices
simultaneously. The flavour changing charged currents are described by the
matrix
\be
V_{CKM} = V_L^U V_L^{D\dagger}\,.
\ee

We now can derive the Yukawa interactions in the Higgs and fermion mass basis.
Using the definitions of Eqs.~(\ref{eq:H2}), (\ref{HiggsDoublets2}), and
(\ref{eq:kappa}-\ref{eq:MasseigenKR}),
the Yukawa interactions (\ref{Yukawa1}) become~(see e.g.~\cite{hep-ph/0504050})
\begin{eqnarray}
\label{Yukawa2}
  -\mathcal{L}_{\rm{Y}}&=&
   \frac{1}{\sqrt{2}}\bar{D}\Bigl[\kd\sba+(\rd P_R+{\rd}^\dagger P_L)\cba \Bigr]Dh
\nonumber\\&&
   \,  +\frac{1}{\sqrt{2}}\bar{D}\Bigl[\kd\cba-(\rd P_R+{\rd}^\dagger P_L)\sba \Bigr]DH
   + \frac{i}{\sqrt{2}}\bar{D}(\rd P_R-{\rd}^\dagger P_L) DA
\nonumber\\&&
 \,+\frac{1}{\sqrt{2}}\bar{U}\Bigl[\ku\sba+(\ru P_R+{\ru}^\dagger P_L)\cba \Bigr]Uh
\nonumber\\&&
    \, +\frac{1}{\sqrt{2}}\bar{U}\Bigl[\ku\cba-(\ru P_R+{\ru}^\dagger P_L)\sba \Bigr]UH
- \frac{i}{\sqrt{2}}\bar{U}(\ru P_R-{\ru}^\dagger P_L) UA 
\nonumber\\
    &&\,+\frac{1}{\sqrt{2}}\bar{L}\Bigl[\kl\sba+(\rl P_R+{\rl}^\dagger P_L)\cba \Bigr]Lh
\nonumber\\&&
    \, +\frac{1}{\sqrt{2}}\bar{L}\Bigl[\kl\cba-(\rl P_R+{\rl}^\dagger P_L)\sba \Bigr]LH
 + \frac{i}{\sqrt{2}}\bar{L}(\rl P_R-{\rl}^\dagger P_L) LA 
 \nonumber\\&&
\nonumber\\&&
    \,+\frac{}{}\Bigl[\bar{U}\bigl(V_{\rm{CKM}} \rd P_R
 -{\ru}^\dagger V_{\rm{CKM}} P_L\bigr)DH^+ + \bar{\nu}\rl P_RL H^+ + \rm{h.c.}\Bigr],
\end{eqnarray}
where $P_{R/L} = (1\pm\gamma_5)/2$.
One can clearly see, that if the Yukawa coupling matrices $\rho^F$ are not
diagonal, there are flavour-changing-neutral-currents (FCNC)
at tree level, which are absent in the Standard Model and are severely constrained by
experiments. Therefore, either
these terms are completely forbidden by certain symmetries or mechanisms, 
or they are sufficiently small to avoid the current experimental bounds. 
An early discussion is the paper by Glashow and Weinberg \cite{Glashow:1976nt}.

\begin{table}
\centering
\begin{tabular*}{0.7\columnwidth}{@{\extracolsep{\fill}}ccccccc}
\hline
Type & $U_R$ & $D_R$ & $L_R$ & $\ru$ & $\rd$ & $\rl$ \\
\hline
I & $+$ & $+$ & $+$ & $\ku\cot\beta$ & $\kd\cot\beta$ & $\kl\cot\beta$ \\
II & $+$ & $-$ & $-$ & $\ku\cot\beta$ & $-\kd\tan\beta$ & $-\kl\tan\beta$ \\
III/Y & $+$ & $-$ & $+$ & $\ku\cot\beta$ & $-\kd\tan\beta$ & $\kl\cot\beta$ \\
IV/X & $+$ & $+$ & $-$ & $\ku\cot\beta$ & $\kd\cot\beta$ & $-\kl\tan\beta$ \\
\hline
\end{tabular*}
\caption{The different types of 2HDM with $Z_2$ symmetry. The nomenclature
follows \cite{Mahmoudi:2009zx}.
The $Z_2$ charges for Higgs doublets are odd or $-1$ for
$\Phi_1$ and even or $+1$ for $\Phi_2$. 
The right-handed fermions have been given different $Z_2$
charges assignment as shown. 
The Yukawa matrices $\rho^F$ are proportional
to the $\kappa^F$ and thus also diagonal with the relation shown in the
last three columns.} \label{tab:Z2}
\end{table}

There are different known solutions to the FCNC problem.
In this paper we study three different cases:
\begin{itemize}

\item \textbf{$Z_2$ symmetry}

If there is only one Higgs doublet coupling to each type
of fermions, the situation becomes the same as in the standard model.
The FCNC couplings vanish completely, known as naturally vanishing FCNC
\cite{Glashow:1976nt}. 
An elegant way to achieve this is to impose a
$Z_2$ symmetry on the Lagrangian and set one of the Higgs doublets and some
of the right handed fermions
to be $Z_2$ odd. The different cases depending on which fermions couple to
the same doublets are listed in Table \ref{tab:Z2}. We also note that
the Higgs sector of the MSSM is
of type II at tree-level.

\item \textbf{Yukawa Alignment}

A more general way to diagonalize the Yukawa matrices simultaneously
is the Yukawa Alignment model \cite{Pich:2009sp}. They proposed that the
Yukawa coupling matrices $\eta^F_1$ and $\eta^F_2$ are proportional to each
other.  So the rotated Yukawa coupling matrices $\kappa^F$ and $\rho^F$
are also proportional to each other and can thus be diagonalized simultaneously.

However, other than the models
with $Z_2$ symmetry, this alignment may be spoiled when going to different energy
scales and some of the non-diagonal couplings leading to FCNC may become
sizable. Studying limits on the proportionality constants
from this source is one of the purposes of the present paper.

\item \textbf{Cheng-Sher Ansatz}

A third possibility is to keep the off-diagonal FCNC elements in the $\rho^F$
naturally small. The best known ansatz of this type was
proposed by Cheng and Sher \cite{Cheng:1987rs}
\begin{eqnarray}
\rho^F_{ij} &=& \lambda^F_{ij} \frac{\sqrt{2m_i m_j}}{v}\,.
\label{CSansatz}
\end{eqnarray}
The $m_i$ are the different fermion masses. Since the diagonal elements
of the $\kappa^F$ have a hierarchy in size corresponding to
the fermion mass hierarchy  it is natural to introduce this also
for the $\rho^F$. The $\lambda^F$ are expected to be
of $\mathcal{O}(1)$ and should be small enough to suppress FCNC to the
observed level. We discuss these limits below.
One should be aware that there are different parameterizations of the Cheng-Sher
ansatz, some papers do not have the factor of $\sqrt{2}$ in
(\ref{CSansatz}), e.g.~\cite{Gupta:2009wn}.

\end{itemize}

\subsection{RGE for Yukawa Couplings in 2HDM}
\label{sec:RGE}

The variation of couplings and masses with the subtraction scale $\mu$
is given by the renormalization group equations (RGE).
The running of Yukawa couplings in the 2HDM can be found in many places, e.g.
\cite{cvetic:1997,Ferreira:2010xe,Branco:2011iw}. We have also rederived them
using the methods of \cite{cvetic:1997}.

Using the notation $\mathcal{D}\equiv 16\pi^2 d/d(\ln\mu)$
the RGEs  for the Yukawa couplings in the general basis are:
\begin{eqnarray}
\mathcal{D} \eta^U_k  & = &
 - A_U \eta^U_k
+ \sum_{\ell=1}^2
\mbox{Tr} \left[ N_c\left(\eta^U_k \eta^{U\dag}_\ell +
 \eta^D_\ell \eta^{D\dag}_k\right)
 + \eta^{L\dag}_k \eta^L_\ell  \right]\eta^U_\ell
\nonumber\\
&&+\frac{1}{2} \sum_{\ell=1}^2 \left[
  \eta^U_\ell \eta^{U\dag}_\ell
+ \eta^D_\ell \eta^{D\dag}_\ell \right] \eta^U_k
+ \eta^U_k \sum_{\ell=1}^2\eta^{U\dag}_\ell \eta^U_\ell
-2 \sum_{\ell=1}^2 \left[\eta^D_\ell \eta^{D\dag}_k\eta^U_\ell\right]\,,
\nonumber
\end{eqnarray}
\begin{eqnarray}
\mathcal{D} \eta^D_k  & = &
- A_D \eta^D_k
+ \sum_{\ell=1}^2
\mbox{Tr} \left[ N_c \left(\eta^D_k \eta^{D\dag}_\ell + \eta^U_\ell \eta^{U\dag}_k\right)
 + \eta^L_k \eta^{L\dag}_\ell   \right]\eta^D_\ell
\nonumber\\
&&+\frac{1}{2} \sum_{\ell=1}^2 \left[\eta^U_\ell \eta^{U\dag}_\ell
+\eta^D_\ell \eta^{D\dag}_\ell \right] \eta^D_k
+ \eta^D_k \sum_{\ell=1}^2\eta^{D\dag}_\ell \eta^D_\ell
-2 \sum_{\ell=1}^2 \left[\eta^U_\ell \eta^{U\dag}_k\eta^D_\ell\right]\,,
\nonumber
\end{eqnarray}
\begin{eqnarray}
\label{eq:RGEyukawaF}
\mathcal{D} \eta^L_k  & = & - A_L \eta^U_k
+ \sum_{\ell=1}^2
\mbox{Tr} \left[ N_c\left(\eta^{U\dag}_k \eta^U_\ell +
 \eta^D_k \eta^{D\dag}_\ell\right)
 + \eta^L_k \eta^{L\dag}_\ell  \right]\eta^L_\ell
\nonumber\\
&&+ \sum_{\ell=1}^2
 \left[ \frac{1}{2}\eta^L_\ell \eta^{L\dag}_\ell\eta^L_k
+  \eta^L_k \eta^{L\dag}_\ell \eta^L_\ell\right]\,.
\end{eqnarray}
where $A_F$ are given by the gauge couplings as follows
\begin{eqnarray}
A_U & = &
3 \frac{(N_c^2 - 1)}
{ N_c } g_3^2 + \frac{9}{4} g_2^2 +
\frac{17}{12} g_1^2 \ ,
\nonumber\\
A_D & = & A_U - g_1^2 \ ,\nonumber\\
A_L &=& {15\over4}g_1^2+{9\over4}g_2^2\,.
\label{eq:AUADAL}
\end{eqnarray}
with $g_1=e/\cos\theta_W$, $g_2=e/\sin\theta_W$, and $g_3=g_s$, $\sin\theta_W$ being the weak mixing angle. In turn
the RGEs for the gauge couplings up to one loop level are
\begin{eqnarray}
\label{eq:RGEgi}
\mathcal{D} (g_1) & = &\left({1\over3}+{10\over9}n_q\right) g_1^3 \,,
\nonumber\\
\mathcal{D} (g_2) & = &- \left(7-{2\over3}n_q\right)g_2^3\,,
\nonumber\\
\mathcal{D} (g_3) & = & -{1\over3}\left(11N_c - 2n_q\right)g_3^3\,.
\end{eqnarray}
$n_q$ is the number of active quarks above energy threshold.
In this paper we will always use $n_q=6$. 
We have checked that using the two-loop running for $g_3$ produces only a
small change in our results.
We thus expect that the effect of running with $n_q=6$
from $m_Z$ to the top threshold rather than $n_q=5$
will not introduce a significant effect. 

Finally the RGEs for the fields and thus for the vacuum expectation
values $e^{\theta_i} v_i$ are:
\begin{eqnarray}
\mathcal{D} (e^{i\theta_k}v_k)&=& -\sum^2_{\ell=1}\mbox{Tr}
\left[N_c
\Big(\eta^U_k \eta^{U\dag}_\ell + \eta^D_\ell\eta^{D\dag}_k\Big) 
 + \eta^L_\ell\eta^{L\dag}_k\right]
e^{i\theta_\ell}v_\ell
\nonumber\\&&
+\left({3\over4}g^2_1 + {9\over4}g^2_2\right)e^{i\theta_k}v_k\,.
\end{eqnarray}

Note that the running of the Yukawa couplings as given in (\ref{eq:RGEyukawaF})
is independent of the couplings in the Higgs potential (\ref{scalarpotential}). They only appear at the two-loop level.
Thus we limit ourselves to studying the evolution of the Yukawa sector by itself and do not include the evolution of the parameters of the Higgs potential. One should keep in mind that the evolution of the latter could also signal the breakdown of a given model. This has for example been studied in~\cite{Ferreira:2009jb} although only including the top Yukawa coupling. A complete one-loop treatment of the Higgs sector would require also the inclusion of the complete Yukawa sector at one-loop. We foresee to include this in future versions of the 2HDMC calculator~\cite{Eriksson:2009ws}.

Using the definitions (\ref{eq:H2}), (\ref{eq:kappa}) and (\ref{eq:rho}),
the RGEs can be rewritten in the Higgs basis.
The vacuum expectation value $v$, the phase difference between the two vacuum
expectation values $\theta$ and the angle $\beta$ relating the general basis and the Higgs
basis satisfy the following  RGEs:
\begin{eqnarray}
\mathcal{D} \left( v^2 \right) & = &
- 2  \mbox{Tr} \left[ N_c\left(
\kappa_0^U \kappa_0^ {U\dag} +
\kappa_0^D \kappa_0^ {D\dag} \right)+
\kappa_0^L \kappa_0^ {L\dag}\right] v^2
+ \left[ \frac{3}{2} g_1^2 + \frac{9}{2} g_2^2 \right] v^2 \,,
\nonumber
\end{eqnarray}
\begin{eqnarray}
\mathcal{D} (\tan \beta) & = &
- \frac{1}{2 \cos^2 \beta}
  \mbox{Tr} \Bigg[N_c\left(
\rho^{U}_0 \kappa_0^{U\dag}+\kappa_0^U \rho^{U\dag}_0
+ \kappa_0^D \rho^{D\dag}_0+ \rho^D_0 \kappa_0^ {D\dag}\right)\nonumber\\
&&\quad+ \kappa_0^L \rho^{L\dag}_0+ \rho^L_0 \kappa_0^ {L\dag}\Bigg]\,,
\nonumber
\end{eqnarray}
\begin{eqnarray}
\label{eq:RGEyukawaHiggs}
\mathcal{D} (\theta) & = &
\frac{1}{i \sin (2 \beta)}
 \mbox{Tr} \Bigg[N_c\left( \kappa_0^U \rho^{U\dag}_0 - \rho^{U}_0 \kappa_0^ {U\dag}\right)
 - N_c\left( \kappa^D_0 \rho^{D\dag}_0- \rho^D_0 \kappa_0^{D\dag}\right)\nonumber\\
&&\quad - \left( \kappa^L_0 \rho^{L\dag}_0- \rho^L_0 \kappa_0^{L\dag}\right)
\Bigg]\,.
\end{eqnarray}
Finally the Yukawa couplings in the Higgs basis, in other words the matrices
$\kappa^F_0$ and $\rho^F_0$ satisfy:
\begin{eqnarray}
\mathcal{D} \left( \kappa_0^U\right) &=&
- A_U \kappa_0^U 
+ \mbox{Tr}\left[ N_c\left( \kappa_0^U \kappa_0^ {U\dag} +
\kappa_0^D \kappa_0^ {D\dag}\right)
+\kappa^{L\dag}_0\kappa^L_0  \right]\kappa_0^U\nonumber\\
&&- \frac{1}{2} \tan \beta  \, \mbox{Tr}\Bigg\{N_c\left(
\kappa_0^U \rho^{U\dag}_0- \rho^{U}_0 \kappa_0^{U\dag} \right)
-N_c\left( \kappa_0^D \rho^{D\dag}_0- \rho^D_0 \kappa_0^{D\dag}  \right)\nonumber\\
&&\quad
-\left(\kappa^L_0\rho^{L\dag}_0 -\rho^L_0\kappa^{L\dag}_0\right)
\Bigg\}\kappa_0^U
\nonumber\\
&& + {\Bigg \{} \frac{1}{2} \left[ \rho^{U}_0 \rho^{U\dag}_0
+ \rho^D_0 \rho^{D\dag}_0 + \kappa_0^U \kappa_0^ {U\dag}
+ \kappa_0^D \kappa_0^ {D\dag} \right] \kappa_0^U +
\kappa_0^ U \left[ \rho^{U\dag}_0 \rho^{U}_0 +
\kappa_0^{U\dag} \kappa_0^U \right]
\nonumber\\
&&\qquad
 - 2 \rho^D_0 \kappa_0^{D\dag} \rho^{U}_0
- 2 \kappa_0^D \kappa_0^ {D\dag} \kappa_0^ U
{\Bigg \}} \,, 
\end{eqnarray}
\begin{eqnarray}
\mathcal{D} \left( \kappa_0^D \right) &=&- A_D \kappa_0^D
+ \mbox{Tr} \left[N_c\left( \kappa_0^U \kappa_0^ {U\dag} +
\kappa_0^D \kappa_0^ {D\dag}\right)
+\kappa^L_0\kappa^{L\dag}_0 \right]\kappa_0^D\nonumber\\
&&
+ \frac{1}{2} \tan\beta \,
\mbox{Tr} \Bigg\{N_c\left(
\kappa_0^U \rho^{U\dag}_0- \rho^{U}_0 \kappa_0^{U\dag} \right)
-N_c\left( \kappa_0^D \rho^{D\dag}_0- \rho^D_0 \kappa_0^{D\dag}  \right)\nonumber\\
&&\quad
-\left(\kappa^L_0 \rho^{L\dag}_0- \rho^L_0 \kappa_0^{L\dag} \right)
\Bigg\} \kappa_0^D
\nonumber\\
&& + {\Bigg \{} \frac{1}{2} \left[ \rho^{U}_0 \rho^{U\dag}_0
+ \rho^D_0 \rho^{D\dag}_0 + \kappa_0^U \kappa_0^ {U\dag}
+ \kappa_0^D \kappa_0^ {D\dag} \right] \kappa_0^D +
\kappa_0^ D \left[ \rho^{D\dag}_0 \rho^D_0 +
\kappa_0^{D\dag} \kappa_0^D \right]
\nonumber\\
&&\quad - 2 \rho^{U}_0 \kappa_0^{U\dag} \rho^D_0
- 2 \kappa_0^U \kappa_0^ {U\dag} \kappa_0^ D
 {\Bigg \}} \,,
\end{eqnarray}
\begin{eqnarray}
 \mathcal{D} \left( \kappa_0^L \right) &=&
-A_L\kappa^L_0
+\mbox{Tr}\Bigg\{N_c \left(\kappa^{U\dag}_0 \kappa^U_0+ \kappa^D_0 \kappa^{D\dag}_0 \right)+
\kappa^{L\dag}_0\kappa^L_0  \Bigg\} {\kappa^L_0}\nonumber\\
&&+{1\over2}\tan\beta \, \mbox{Tr} \Bigg\{N_c\left(
\kappa_0^U \rho^{U\dag}_0- \rho^{U}_0 \kappa_0^{U\dag} \right)
-N_c\left( \kappa_0^D \rho^{D\dag}_0- \rho^D_0 \kappa_0^{D\dag}  \right)\nonumber\\
&&\quad - \left( \kappa^L_0 \rho^{L\dag}_0- \rho^L_0 \kappa_0^{L\dag}\right)\Bigg\}\kappa^L_0\nonumber\\
 &&+\frac{1}{2}\left(\rho^L_0\rho^{L\dag}_0+\kappa_0^L\kappa_0^{ L\dag}\right)\kappa^L_0
+\kappa^L_0\left(\rho^{L\dag}_0\rho^L_0+\kappa_0^{\dag L}\kappa_0^{ L}\right)\,,
\end{eqnarray}
\begin{eqnarray}
\mathcal{D} (\rho^{U}_0) &=&- A_U \rho^{U}_0
+ 2 \mbox{Tr} \left[ N_c\left(\rho^{U}_0 \kappa_0^{U\dag}
+ \kappa_0^D \rho^{D\dag}_0\right)
+\kappa^L_0\rho^{L\dag}_0\right] \kappa_0^ U\nonumber\\
&&+ \mbox{Tr} \left[ N_c\left(\rho^{U}_0 \rho^{U\dag}_0 +
\rho^D_0 \rho^{U\dag}_0 \right)
+ \rho^L_0\rho^{L\dag}_0\right] \rho^{U}_0
\nonumber\\
&&
- \frac{1}{2}\cot \beta \,
\mbox{Tr} \Bigg\{N_c \left(
\kappa_0^U \rho^{U\dag}_0- \rho^{U}_0 \kappa_0^{U\dag} \right)
-N_c\left( \kappa_0^D \rho^{D\dag}_0- \rho^D_0 \kappa_0^{D\dag}  \right)\nonumber\\
&&\quad-\left(\kappa^L_0\rho^{L\dag}_0 -\rho^L_0\kappa^{L\dag}_0\right)\Bigg\} \rho^{U}_0
\nonumber\\
&& + \Bigg \{ \frac{1}{2} \left[ \rho^{U}_0 \rho^{U\dag}_0
+ \rho^D_0 \rho^{D\dag}_0 + \kappa_0^U \kappa_0^ {U\dag}
+ \kappa_0^D \kappa_0^ {D\dag} \right] \rho^{U}_0 +
\rho^{U}_0 \left[ \rho^{U\dag}_0 \rho^{U}_0 +
\kappa_0^{U\dag} \kappa_0^U \right]
\nonumber\\
&& \quad- 2 \rho^D_0 \rho^{D\dag}_0 \rho^{U}_0
- 2 \kappa_0^D \rho^{D\dag}_0 \kappa_0^ U
\Bigg \}\,,
\end{eqnarray}
\begin{eqnarray}
\mathcal{D}(\rho^D_0) &=&- A_D \rho^D_0 
+2 \mbox{Tr} \left[ N_c\left(
\kappa_0^U \rho^{U\dag}_0 + \rho^D_0 \kappa_0^{D\dag}\right)
+\rho^L_0 \kappa^{L\dag}_0 \right] \kappa_0^ D\nonumber\\
&&+ \mbox{Tr} \left[N_c\left( \rho^{U}_0 \rho^{U\dag}_0 +
\rho^D_0 \rho^{D\dag}_0
+\rho^L_0\rho^{L\dag}_0 \right) \right] \rho^D_0  \nonumber\\
&& +\frac{1}{2} \cot \beta  \, 
\mbox{Tr} \Bigg\{ N_c\left(
\kappa_0^U \rho^{U\dag}_0- \rho^{U}_0 \kappa_0^{U\dag} \right)
-N_c\left( \kappa_0^D \rho^{D\dag}_0- \rho^D_0 \kappa_0^{D\dag}  \right)\nonumber\\
&&\quad
 -\left(\kappa^L_0 \rho^{L\dag}_0- \rho^L_0 \kappa_0^{L\dag}\right)
   \Bigg\}\rho^D_0
\nonumber\\
&& + {\Bigg \{} \frac{1}{2} \left[ \rho^{U}_0 \rho^{U\dag}_0
+ \rho^D_0 \rho^{D\dag}_0 + \kappa_0^U \kappa_0^ {U\dag}
+ \kappa_0^D \kappa_0^ {D\dag} \right] \rho^D_0 +
\rho^D_0 \left[ \rho^{D\dag}_0 \rho^D_0 +
\kappa_0^{D\dag} \kappa_0^D \right]
\nonumber\\
&& - 2 \rho^{U}_0 \rho^{U\dag}_0 \rho^D_0
- 2 \kappa_0^U \rho^{U\dag}_0 \kappa_0^ D
{\Bigg \}}\,,
\end{eqnarray}
\begin{eqnarray}
\mathcal{D}\rho^L_0  &=&-A_L\rho^L_0
+2\mbox{Tr}\Bigg\{N_c \left(\kappa^U_0\rho^{U\dag}_0 + \rho^D_0 \kappa^{D\dag}_0 \right)+
\rho^L_0 \kappa^{L\dag}_0  \Bigg\} \kappa^L_0\nonumber\\
&&
+\mbox{Tr}\Bigg\{N_c \left( \rho^{U}_0\rho^{U\dag}_0+ \rho^D_0 \rho^{D\dag}_0\right)+
\rho^L_0 \rho^{L\dag}_0 \Bigg\} \rho^L_0\nonumber\\
&&+{1\over2}\cot\beta  \, \mbox{Tr} \Bigg\{N_c\left(
\kappa_0^U \rho^{U\dag}_0- \rho^{U}_0 \kappa_0^{U\dag} \right)
-N_c\left( \kappa_0^D \rho^{D\dag}_0- \rho^D_0 \kappa_0^{D\dag}  \right)\nonumber\\
&&\quad
 - \left( \kappa^L_0 \rho^{L\dag}_0- \rho^L_0 \kappa_0^{L\dag}\right)\Bigg\}\rho^L_0\nonumber\\
 &&+\frac{1}{2}\left(\rho^L_0\rho^{L\dag}_0+\kappa_0^L\kappa_0^{ L\dag}\right)\rho^L_0
+\rho^L_0\left(\rho^{L\dag}_0\rho^L_0+\kappa_0^{\dag L}\kappa_0^{ L}\right)\,.
\end{eqnarray}

Before ending this section we note that the $\tan\beta$ dependent terms in the evolution equations for the Yukawa couplings disappear in the real case. In the CP-violating case $\rho$ is no longer basis-independent and therefore there is a residual dependence on $\tan\beta$ in this case.  For a thorough discussion of basis independent quantities in the CP-violating case we refer to~\cite{hep-ph/0602242}.

\section{Constraints and SM input }
\label{sec:constraints}

\subsection{Low-energy constraints on $\lambda^F_{ij}$ }
In the recent review of 2HDM \cite{Branco:2011iw}, the authors have
given a comprehensive overview on the latest constraints on the
$\lambda^F_{ij}$. The most stringent ones are in the quark sector, coming from the neutral meson mixing, and we will therefore limit ourselves to these constraints in the following.

The master formula for $F^0-\bar F^0$ mixing mediated by tree level Higgs scalars in the 
vacuum insertion approximation can be found in \cite{Atwood:1996vj}:
\begin{eqnarray}
\label{NeutralMixing}
\Delta M_F &=& {|\rho^F_{ij}|^2 \over M_F}
  \left[S_F\left({\cba^2\over m^2_h} + {\sba^2\over m^2_H}\right) + {P_F\over m^2_A} \right]\\
S_F &=& {1\over6}B_F f^2_F M^2_F\left[1+\frac{M^2_F}{(m_i+m_j)^2} \right]\nonumber\\
P_F &=& {1\over6}B_F f^2_F M^2_F\left[1+\frac{11M^2_F}{(m_i+m_j)^2} \right]\nonumber
\end{eqnarray}
Here $M_F$ and $\Delta M_F$ are the mass and mass difference of the
neutral mesons respectively, and $f_F$ is the corresponding pseudo-scalar decay constant. The
parameter $B_F$ is defined as the ratio of the actual matrix element compared to its value in the vacuum insertion approximation~\cite{Atwood:1996vj}. The numerical values of the parameters we use are listed in Table~\ref{table:mesons}.

\begin{table}
\begin{center}
\begin{tabular}{cccc}
\hline
Meson                 & $M_F$(GeV)      & $B_F$                              & $f_F$ (GeV)                           \\
\hline
$K^0\,( d\bar s)$    &0.4976~\cite{pdg} &$0.75\pm 0.026$~\cite{Laiho:2009eu} &$0.1558\pm0.0017$~\cite{Laiho:2009eu} \\
$D^0\,(\bar uc)$     &1.8648~\cite{pdg} &$0.82\pm0.01$ ~\cite{Lunghi:2007ak} & 0.165  ~\cite{Lunghi:2007ak}        \\
$B_d^0\, (d \bar b)$ & 5.2795~\cite{pdg}&$1.26\pm0.11$~\cite{Laiho:2009eu}   &$0.1928\pm0.0099$ ~\cite{Laiho:2009eu}\\
$B_s^0\, ( s\bar b)$ & 5.3663~\cite{pdg}&$1.33\pm0.06$~\cite{Laiho:2009eu}   &$0.2388\pm0.0095$ ~\cite{Laiho:2009eu}\\
\hline
\end{tabular}
\caption{Parameters of the neutral mesons $K^0$, $D^0$,$B^0_d$ and $B^0_s$.}
\label{table:mesons}
\end{center}
\end{table}

To calculate the limits on $\lambda^F_{ij}$, we require that the sum of the
SM and 2HDM theoretical predictions for $\Delta M_F$ does not exceed
the experimental value by more than 2 standard deviations:
\begin{eqnarray}
\label{eq:limit}
\Delta M_F^{\mathrm{SM}} + \Delta M_F^{\mathrm{2HDM}}
\leq \Delta M_F^{\mathrm{expt}} + 2\sigma
\end{eqnarray}
where $\sigma = \sqrt{\sigma^2_{\mathrm{expt}}+\sigma^2_{\mathrm{SM}}}$ is a combination of the experimental and theoretical uncertainties. 
For the $K^0 - \bar K^0$ and $D^0 - \bar D^0$ mixing, the non-perturbative interactions make the SM calculation very difficult. Here we therefore simply assume that the 2HDM contribution is not larger than the experimental value by more than 2 standard deviations. This corresponds to setting the SM contribution to zero in Eq.~(\ref{eq:limit}) as was done in~\cite{Gupta:2009wn}. The experimental and SM values we thus use are listed below.
\begin{enumerate}
  \item $K^0-\bar K^0$:
  
  \begin{center}
  \begin{tabular}{rcl}
  $\Delta M_{K^0}^{\rm expt}$ & = & $(3.483\pm 0.006)\times 10^{-15}$  GeV \cite{pdg}\\
  $\Delta M^{\rm SM}_{K^0}$ & = & 0
  \end{tabular}
  \end{center}
  
  \item $D^0-\bar D^0$:
  
  \begin{center}
  \begin{tabular}{rcl}
    $\Delta M^{\rm expt}_{D^0}$ &=& $1.57^{+0.39}_{-0.415}\times10^{-14}$  GeV  \cite{pdg}\\
    $\Delta M^{\rm SM}_{D^0}$ &=& 0
  \end{tabular}
  \end{center}

  \item $B^0_d-\bar B^0_d$:
  
  \begin{center}
  \begin{tabular}{rcl}
 $\Delta M^{\rm expt}_{B_d}$&=& $(3.344 \pm 0.0197 \pm  0.0197) \times 10^{-13}$  GeV \cite{pdg}\\
  $\Delta M^{\rm SM}_{B_d}$ &=& $3.653^{+0.48} _{-0.30} \times 10^{-13}$  GeV  \cite{Lenz:2010gu}
  \end{tabular}
  \end{center}

  \item $B^0_s-\bar B^0_s$:
  
  \begin{center}
  \begin{tabular}{rcl}
$\Delta M^{\rm expt}_{B_s}$  &=&$(116.668\pm0.270\pm0.171)\times10^{-13}$  GeV  \cite{LHCb}\\
$\Delta M^{\rm SM}_{B_s}$ &=& $110.6^{+17.1}_{-9.9}\times10^{-13}$  GeV  \cite{Lenz:2010gu}\\
  \end{tabular}
  \end{center}
\end{enumerate}

The 2HDM contribution is then calculated using   Eq.~(\ref{NeutralMixing}). We note that 
the quark masses appearing in Eq.~(\ref{NeutralMixing}) are the low energy ones defined more or less at the scale of the respective meson masses. For internal consistency we use the following values from ref. \cite{Xing:2007fb} (in GeV):
\begin{eqnarray*}
m_u(2 \mbox{ GeV}) &=& 2.2\times10^{-3}\,,\ \ \quad m_c(m_c) = 1.25\,; \\
m_d(2 \mbox{ GeV}) &=& 5.0\times 10^{-3}\,,\ \ \quad m_s(2 \mbox{ GeV}) = 0.095\,, \quad m_b(m_b) = 4.2\,. 
\end{eqnarray*}
However, the impact of the actual quark masses used is very small since the masses appearing in $|\rho^F_{ij}|^2$ and the dominant pseudo-scalar matrix element ${M^2_F}/{(m_i+m_j)^2}$ essentially cancel, and we get similar results using the masses defined at $m_Z$ instead.

From  Eq.~(\ref{NeutralMixing}) we can see that the main uncertainty of this estimate is due to the unknown masses of the CP-even  and CP-odd Higgs bosons. It is also clear that the contribution to the mixing from the CP-odd exchange is much larger due to the extra factor 11 in $P_F$ for the dominant pseudo-scalar matrix element. 
We will consider three different representative cases.  
We also remind the reader that in some cases there is  no factor of $\sqrt{2}$ in the definition of  $\lambda^F_{ij}$. 
With all this in mind we get the following constraints on $\lambda^F_{ij}$:

\begin{itemize}
  \item $m_h = m_H = m_A = 120$ GeV
   \begin{eqnarray*}
   \lambda_{uc} &\lesssim& 0.13\,,\\
   \lambda_{ds} &\lesssim& 0.08,\quad \lambda_{db} \lesssim 0.03,\quad \lambda_{sb} \lesssim 0.05\,.
   \end{eqnarray*}

  \item $m_h = m_H = m_A = 400$ GeV
     \begin{eqnarray*}
   \lambda_{uc} &\lesssim& 0.44\,, \\
   \lambda_{ds} &\lesssim& 0.27\, ,\quad \lambda_{db} \lesssim 0.12\, ,\quad \lambda_{sb} \lesssim 0.18\,.
   \end{eqnarray*}

  \item $m_h = m_H = 120$ GeV  $m_A = 400$ GeV
   \begin{eqnarray*}
   \lambda_{uc} &\lesssim& 0.30 \,,\\
   \lambda_{ds} &\lesssim& 0.20 \,,\quad \lambda_{db} \lesssim 0.08 \,,\quad \lambda_{sb} \lesssim  0.12 \,.
   \end{eqnarray*}
\end{itemize}

The first and second cases are examples of typical low and intermediate masses for the Higgs bosons, whereas  the last case illustrates that the main restriction comes from the exchange of the CP-odd Higgs. All in all we conclude from these different cases that a representative value for these constraints is given by $\lambda^F_{i \neq j} \lesssim 0.1$ and this is the generic value we will use when analyzing the effects of $Z_2$ breaking in the running of the Yukawa couplings in the next section.

\subsection{General input}

For the RGE evolution towards high scales we need
a set of input parameters at the low scale $\mu=m_Z= 91.186$ GeV.
The experimental input
we have are the masses and the measured parameters of the
CKM-mixing matrix as well as the gauge couplings. 
We have neglected constraints coming from the neutrino
sector. The quark and charged lepton masses at the scale $m_Z$
we take from Ref.~\cite{Xing:2007fb}, their
values  are (in GeV)
\begin{eqnarray*}
m_u &=& 1.29\times10^{-3}\,,\ \ \quad m_c = 0.619\,, \quad m_t = 171.7\,;\\
m_d &=&2.93\times 10^{-3}\,,\ \ \quad m_s = 0.055\,, \quad m_b = 2.89\,; \\
m_e &=& 0.487\times10^{-3}\,,\quad m_\mu = 0.103\,,\quad m_\tau = 1.746\,.
\end{eqnarray*}
For the $3\times3$ CKM matrix we use the PDG  \cite{pdg} phase convention
\be
V_{CKM}= \left(\begin{array}{ccc}
c_{12}c_{13}  & s_{12}c_{13} & s_{13}e^{-i\delta}\\
-s_{12}c_{23}-c_{12}s_{23}s_{13}e^{i\delta} &
   c_{12}c_{23}-s_{12}s_{23}s_{13}e^{i\delta} & s_{23}c_{13}\\
s_{12}s_{23}-c_{12}c_{23}s_{13}e^{i\delta} &
    -c_{12}s_{23}-s_{12}c_{23}s_{13}e^{i\delta} & c_{23}c_{13}
\end{array}\right)\,,
\ee
where $s_{ij}=\sin\theta_{ij}$ and $c_{ij}=\cos\theta_{ij}$.
We will also use this convention for the phases at the high scale.
The values for the angles and the phase follow from \cite {pdg}
\begin{eqnarray}
s_{21}&=&\lambda\,,\quad s_{23} = A\lambda^2\,,
\nonumber\\
s_{13}e^{i\delta} &=&
\frac{A\lambda^3\left(\bar\rho+i\bar\eta\right)\sqrt{1-A^2\lambda^4}}
{\sqrt{1-\lambda^2}\left[1-A^2\lambda^4\left(\bar\rho+i\bar\eta\right)\right]}
\,.
\end{eqnarray}
with
\begin{equation}
\lambda=0.2253\,,\quad A=0.808\,,\quad\bar\rho=0.132\,,\quad
\bar\eta=0.341\,.
\end{equation}

There is of course still a large freedom in how one chooses the remaining freedom
at the weak scale $m_Z$. We chose to put the CKM-mixing always in
the down quark sector and have thus at the EW scale
\begin{eqnarray*}
V_L^{U} &=& V_R^{U} = I\\
V_L^{D} &=& V^\dag_{CKM} \qquad V_R^{D} = I\\
V_L^L &=& V_R^L = I\,.
\label{qmass}
\end{eqnarray*}
The last two are a consequence of our neglecting neutrino masses and mixings.
The Yukawa couplings at the EW scale are thus:
\begin{eqnarray*}
(\kappa^U_0)_{ij} &=& \kappa^U_{ij} =  \frac{\sqrt2m_i}{v}\,,\qquad \ \qquad\quad
(\rho^U_0)_{ij} = \rho^U_{ij} \qquad \qquad(i,j = u,c,t)\\
(\kappa^D_0)_{ij} &=& V_{\rm CKM}\,\kappa^D_{ij} =V_{\rm CKM}\frac{\sqrt2m_i}{v}\,,\ \
(\rho^D_0)_{ij} = V_{\rm CKM}\,\rho^D_{ij}  \quad (i,j = d,s,b)\\
(\kappa^L_0)_{ij} &=& \kappa^L_{ij}  = \frac{\sqrt2m_i}{v}\,,\qquad \ \qquad\quad
(\rho^L_0)_{ij} = \rho^L_{ij} \qquad\qquad (i,j = e,\mu,\tau)
\end{eqnarray*}

At any energy higher than the EW scale, the Yukawa couplings $\kappa_0$ and $\rho_0$ in general become non-diagonal and complex. Thus they need to be transformed to the mass eigenstates by the bi-diagonalization defined in 
Eq.~(\ref{eq:MasseigenKR}) in order to give $\kappa$ and $\rho$. The latter can then be used together with the diagonal elements of the former to calculate $\lambda^F_{i \neq j}$. When performing the bi-diagonalization we always keep to the PDG conventions for how to write the CKM matrix.

For the electroweak VEV we use $v^2=1/(\sqrt{2}G_F)$ with $G_F=1.16637\cdot10^{-5}$ GeV$^{-2}$ from PDG~\cite{pdg} and for the phase difference between the two VEVs we start from $\theta=0$ such that there is no spontaneous CP-violation.
For the gauge couplings we use the PDG~\cite{pdg} values:
$\alpha=1/127.91$, $\alpha_s=0.118$ and for the weak mixing angle we use the on-shell value $\sin^2\theta_W=0.2233$. 

\section{RGE analysis}
\label{sec:analysis}

We have implemented the RGE equations in the Higgs basis given above in three different computer codes. The matrix operations have been performed with either the C++ template library {\it Eigen} \cite{eigen} or the {\it GNU Scientific Library (GSL)}\cite{gsl} and the in total 114 ordinary differential equations are handled by the ODE-solver in GSL using the explicit Runge-Kutta-Fehlberg (4,5) method.
The programs have been tested against each other and also by comparing with the results from~\cite{cvetic:1997}.

In this section we will start by briefly exploring the behavior of $Z_2$-symmetric models and then study a number of $Z_2$-breaking models in more detail.

\subsection{${Z}_2$-symmetric models}

From table \ref{tab:Z2} and the definitions of $\kappa^F$ and
$\rho^F$, we get the diagonal elements of $\lambda^F_{ii}$ in terms of $\tan\beta$
for the four different 2HDM types as shown in table \ref{tab:Z2lambda}. 
Since in this case the Yukawa couplings are given by $\tan\beta$ it is a real physical parameter. 
In addition the evolution of the Yukawa couplings will only depend on the initial value of $\tan\beta$. 

Since the $Z_2$-symmetry is enforced the Yukawa couplings stay diagonal and the only thing that can happen during the evolution is that one or more of the Yukawas will blow up due to the presence of a Landau pole. This signals the breakdown of the perturbative description and calls for a new theory at the corresponding energy scale. The position of the Landau pole will depend on the initial value of $\tan\beta$ and which of the four types we are considering. 

\begin{table}
\centering
\begin{tabular*}{0.7\columnwidth}{@{\extracolsep{\fill}}cccc}
\hline
Type &  $\lambda^U_{ii}$ & $\lambda^D_{ii}$ & $\lambda^L_{ii}$ \\
\hline
I     &  $1/\tan\beta$  & $ 1/\tan\beta$ & $1/\tan\beta$\\
II    &  $1/\tan\beta$  & $ -\tan\beta$        & $-\tan\beta$ \\
III/Y &  $1/\tan\beta$  & $-\tan\beta$         & $1/\tan\beta$ \\
IV/X  &  $1/\tan\beta$  & $ 1/\tan\beta$ & $-\tan\beta$ \\
\hline
\end{tabular*}
\caption{
The diagonal $\lambda^F_{ii}$ in  2HDM models with $Z_2$ symmetry.}
\label{tab:Z2lambda}
\end{table}

\begin{figure}
\centering
\includegraphics[width=14cm, viewport=50 115 800 460,clip]{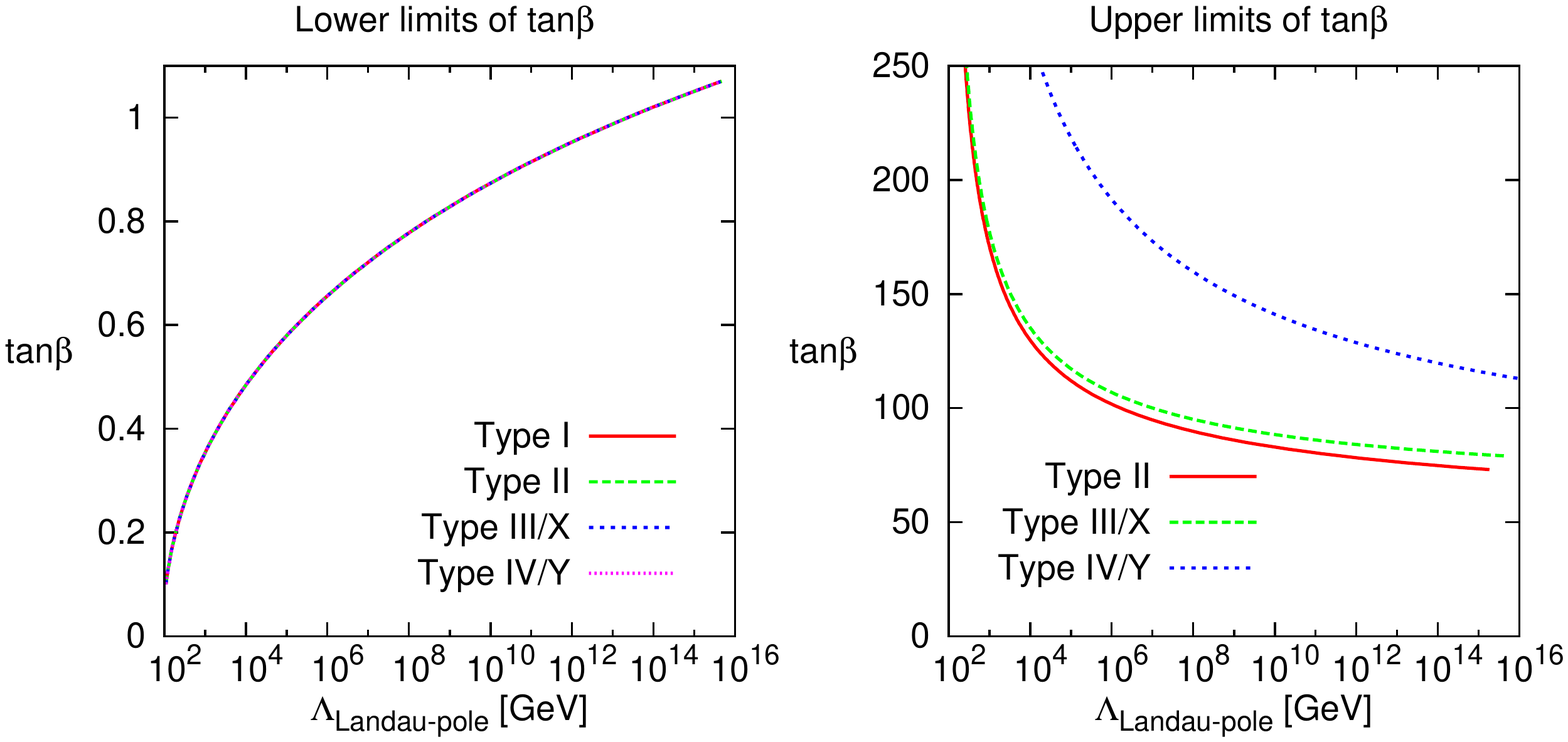}
\caption{The starting value of $\tan\beta$ as a function of the position of the corresponding Landau pole ($\Lambda_{\rm Landau-pole}$) in the different 2HDM types. There are lower limits on $\tan\beta$ for all
    four types (left), but only type II, type III/X and type IV/Y have an upper limit of $\tan\beta$ (right).}
    \label{plot:LPtypes}
\end{figure}

In Fig. \ref{plot:LPtypes} we show the position of the Landau pole
as a function of the input $\tan\beta$.
For the lower limits, the results are almost the
same for all types, and the lines are more or less on top of each other. This is natural since in this regime the evolution is essentially driven by $\lambda_{tt}$, which is the same in all types. For the upper limits, on the other hand, there are some differences. First of all there is no upper limit on $\tan\beta$ in the type I model, which means there is no Landau pole
below $10^{16}$ GeV if the input $\tan\beta>1.1$. For the other types, the upper limits are shown in the right panel of
Fig. \ref{plot:LPtypes}. The differences can be understood from whether the evolution is driven by $\lambda_{bb}$ (type III), $\lambda_{\tau\tau}$ (type IV) or both (type II).

\subsection{$Z_2$-breaking models}

Before starting to analyze the $Z_2$-breaking models we note that, as shown by \cite{Ferreira:2010xe}, 
the $Z_2$-symmetry of the RGE's is still preserved if all the $\lambda^F$'s for the different types are
rescaled with a factor $x$ for the $\cot\beta$ ones and $1/x$ for the $\tan\beta$ ones. In addition, when  $\tan\beta$ is no longer related to the Yukawa couplings it does not have any physical meaning, since it only reflects the basis choice for the general 2HDM. In the following we will only be considering cases with $\rho$ real at the starting scale. This means that the only source of CP-violation is from the CKM-matrix. Thus the CP-violating effects will be small and therefore the dependence on $\tan\beta$ very limited. We have verified this numerically for a number of cases and in the following we set $\tan\beta=1$.

In this subsection, we will also explore the non-diagonal elements of FCNC Yukawa couplings.
We know that in the $Z_2$ symmetric case, the tree level FCNC couplings
will remain equal to zero (up to the numerical precision) up to arbitrarily high energy scales since they are protected by the symmetry.
However once we break the $Z_2$ symmetry in some way, this protection is not effective
anymore and the off-diagonal elements $\lambda^F_{i \neq j}$ may start to grow.

The actual values of the non-diagonal FCNC Yukawa couplings
$\rho^F_{i \neq j}$ at different energy scales will depend on how much we break the $Z_2$ symmetry.
We can thus use the size of the $\lambda^F_{i \neq j}$ as a measure of how severe different types of $Z_2$ symmetry breaking are.
Of course we do not know how large the $\lambda^F_{i \neq j}$ can be at higher scales. Still it is reasonable to assume that the values should not be widely different from the ones at the EW scale. Thus we will use a generic value of $\lambda^F_{i \neq j} \leq 0.1$ as a limit on how much  $Z_2$ symmetry breaking should be allowed and see at which energy scale this limit is reached. 

The argument behind this is essentially that  we can use the RGE evolution to analyze the stability of the assumptions underlying different 2HDMs under variations of the scale where the model is defined. A large sensitivity indicates 
that the assumptions behind the model are not stable meaning that they are either fine-tuned or incomplete such that there for example will be additional particles appearing when going to a higher energy. 
From this respect we will thus study both the appearance of a Landau pole as well as off-diagonal Yukawa couplings leading to FCNC larger than experimentally allowed at the EW scale. 
We also note that as will become clear below there is a small dependence on at which scale we apply the above argument. Requiring stability up to $10^3$ GeV gives very similar constraints on the amount of $Z_2$-breaking that is allowed as when using  $10^{15}$ GeV. 

As an alternative way of assessing the amount of $Z_2$-breaking that is allowed by the experimental constraints from FCNC at the EW scale we have also considered a set-up where the $Z_2$-symmetry is broken at a high scale.
In this set-up we start from a $Z_2$-symmetric model at the EW scale and then evolve it to the high scale of interest. The resulting model is then used as the starting point for exploring different ways of breaking the $Z_2$ symmetry. Once the $Z_2$-breaking has been introduced the different models are then evolved down to the EW scale for comparison with the experimental constraints.  We have verified that in representative cases the results obtained in this way are very similar to the first approach and therefore we will not go into any more details.

There are many possibilities to break the $Z_2$ symmetry and in the following we will consider three ways: aligned, diagonal and non-diagonal $\lambda^F_{ij}$ as defined below. In most cases we will concentrate on the effects of breaking the symmetry starting from a type I or type II model. The reasons for this is on the one hand that these models are the most  well
studied cases in the literature and on the other hand that it is in the quark sector that we have the most stringent constraints on the FCNC Yukawa couplings. Thus the breaking of the $Z_2$ symmetry in the lepton sector will typically have small effects.

In order to be able to separate the effects of breaking the $Z_2$ symmetry in different ways we will limits ourselves to breaking the symmetry in one specific way at a time.

We start by noting that in the $Z_2$ symmetric models at least two of the $\lambda^F$ are always equal whereas the third one is the same as the other two in type I and the negative inverse of them in the other types. When going to the aligned models we will therefore keep two of the sectors in fulfillment with the $Z_2$-symmetry and only break the symmetry through the relation to the third sector. In other words either setting 
$\lambda^D_{ii}  = \lambda^L_{ii}$,  $\lambda^U_{ii}  = \lambda^L_{ii}$, or  $\lambda^U_{ii}  = \lambda^D_{ii}$ and letting $\lambda^F_{ii}$ of the third sector vary independently of the other two.

Another way of breaking the $Z_2$-symmetry is by keeping the  $ \lambda^F_{ij}$ diagonal but letting the individual diagonal elements be non-equal as has been studied by  Mahmoudi and St{\aa}l \cite{Mahmoudi:2009zx}. We will analyze the effects of this type of breaking in the up and down sectors separately again starting from the $Z_2$-symmetric cases with either $\lambda^D_{ii}  = \lambda^L_{ii}  = \lambda_{tt}$ or $\lambda^D_{ii}  = \lambda^L_{ii}  =-{1 / \lambda_{tt}}$. In other words using the type I or II $Z_2$-symmetries as starting point.

The third way of breaking the $Z_2$-symmetry that we will consider is by setting the non-diagonal elements of $\lambda^F_{ij}$ nonzero already at the starting scale. Again we will consider setting the up-sector and down-sector non-diagonal elements non-zero separately and apply the type I or type II symmetries for the diagonal elements.

\subsubsection{Aligned models}
We start by analyzing the three different versions of Aligned models with $\lambda^U$, $\lambda^D$, and $\lambda^L$ pairwise equal. Based on the similarities with the $Z_2$-symmetric models we call them I/II, III, and IV respectively and their free parameters are as follows
\begin{itemize}
  \item Aligned I/II:    $ \lambda^U_{ii} , \quad \lambda^D_{ii}  = \lambda^L_{ii}$
  \item Aligned III:    $ \lambda^D_{ii} , \quad \lambda^U_{ii}  = \lambda^L_{ii}$
  \item Aligned IV:    $ \lambda^L_{ii} , \quad \lambda^U_{ii}  = \lambda^D_{ii}$
\end{itemize}

\begin{figure}
\includegraphics[width=6.2cm]{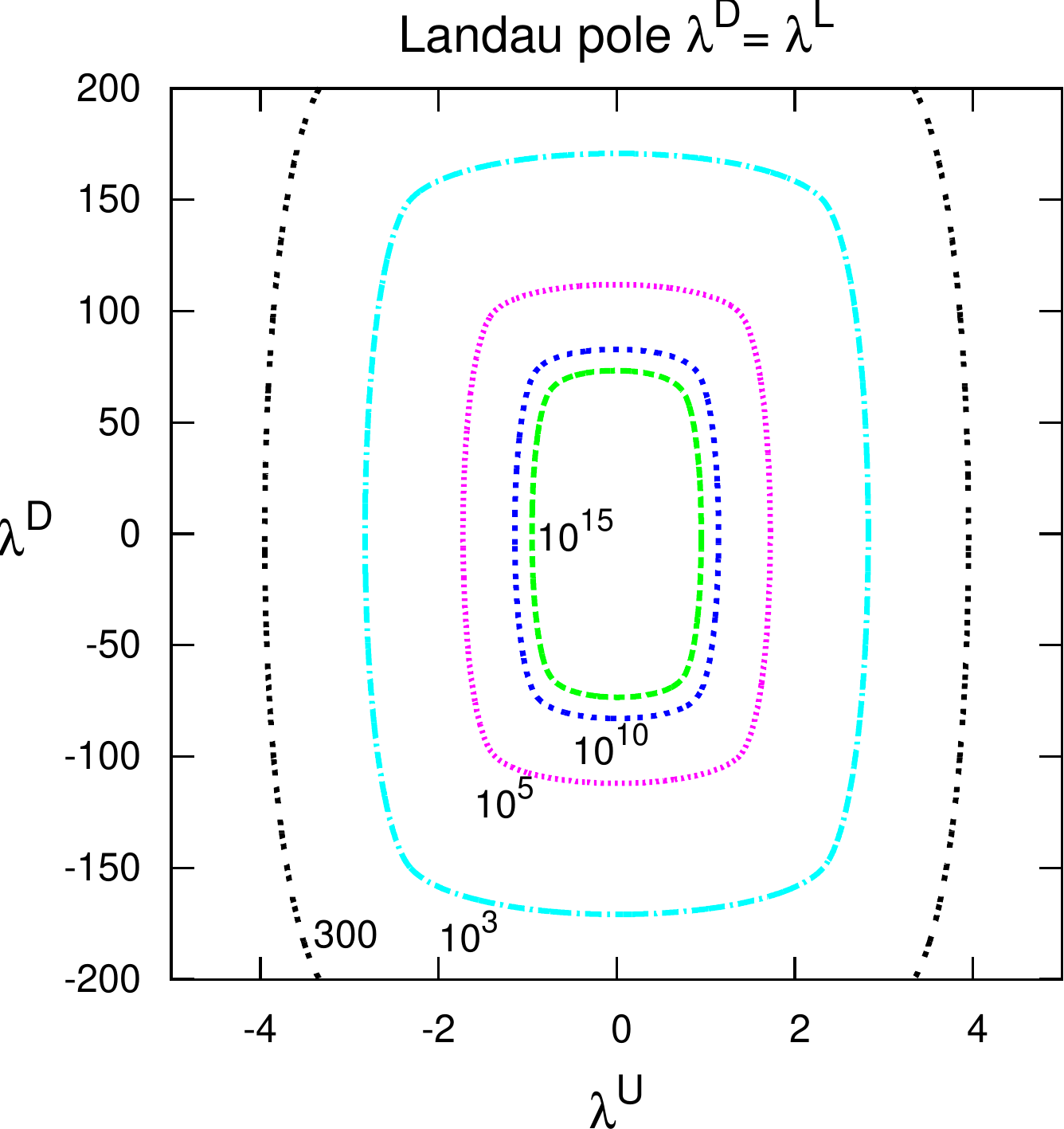} \hspace*{5mm}
\includegraphics[width=6.2cm]{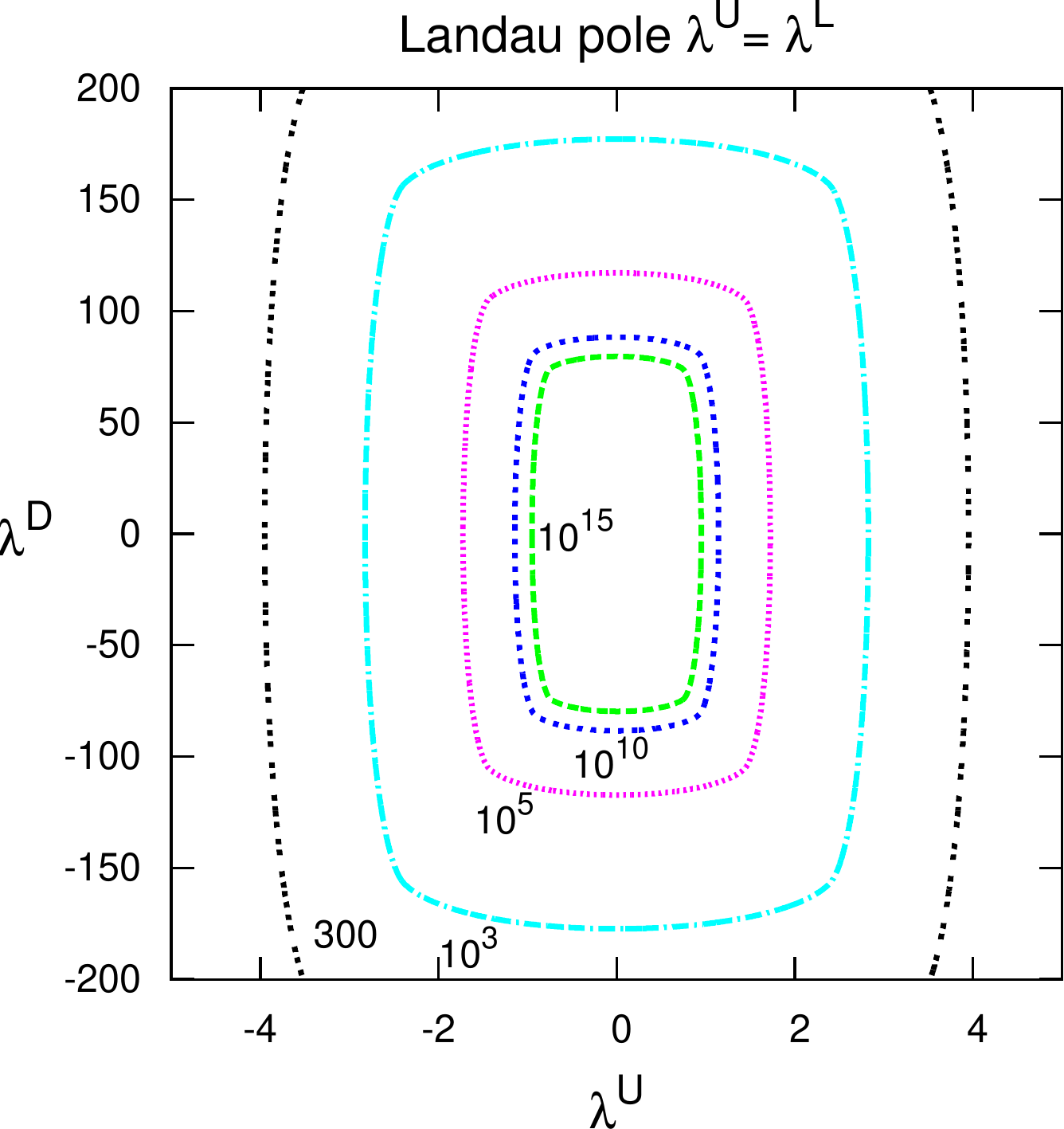} \newline \\
\includegraphics[width=6.2cm]{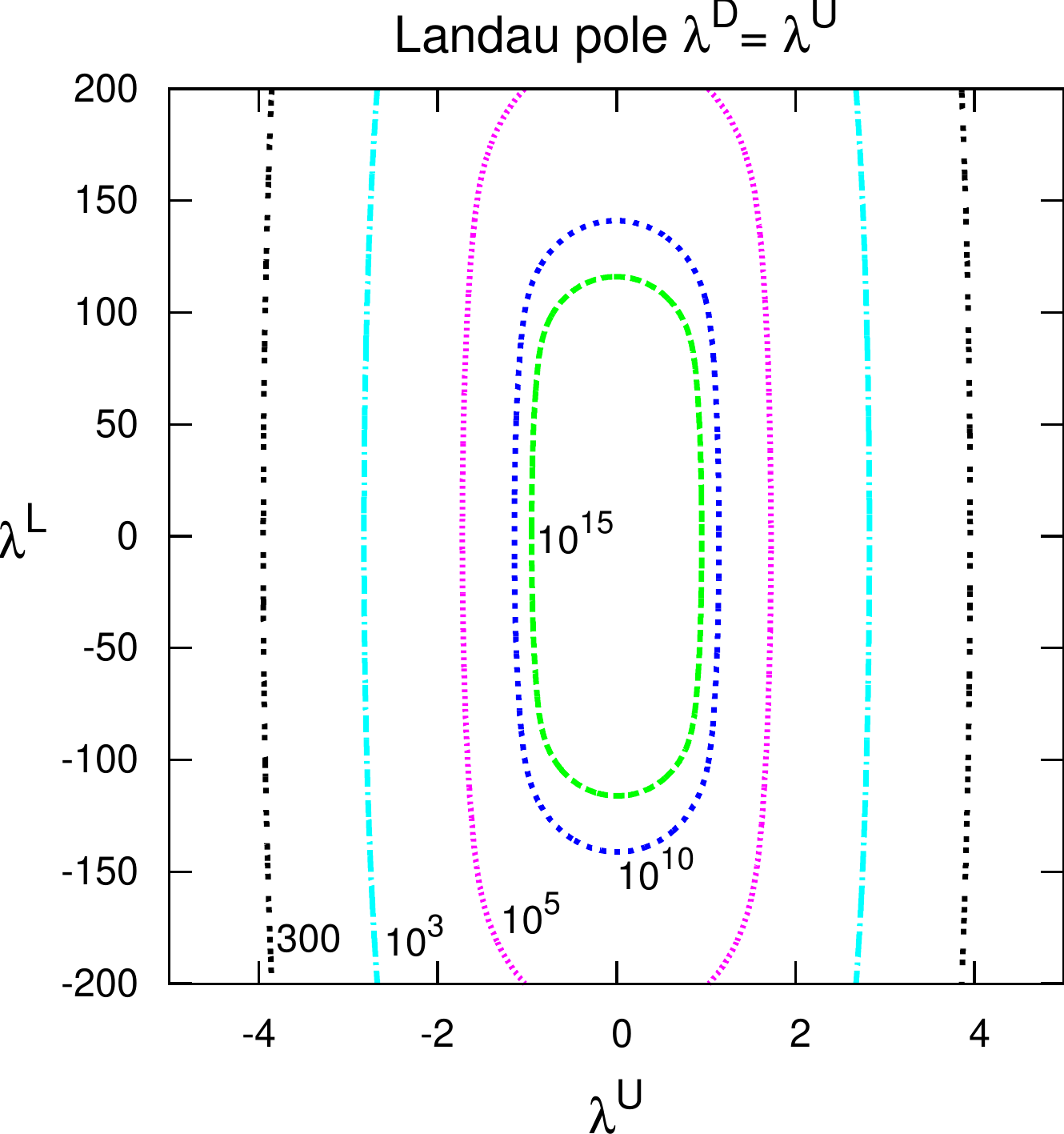}
  \caption{The energy scale at which the Landau pole is encountered as a function of pairwise combinations of the starting values for $\lambda^U_{ii}$, $\lambda^D_{ii}$, and $\lambda^L_{ii}$ as indicated in the figure for the three different versions of aligned models explained in the text. The areas inside a given contour are allowed by the requirement of not having a Landau pole. The different contours are as follows starting from the center: $10^{15}$, $10^{10}$, $10^{5}$, $10^{3}$, and $300$ GeV.}
\label{plot:2DA_landau}
\end{figure}

First we consider the effects of requiring that there is no Landau pole encountered when evolving to higher scales.
We therefore plot in  Fig.~\ref{plot:2DA_landau} the scale at which the Landau pole is reached as a function of
the starting values for pairs of $\lambda^U$, $\lambda^D$, and $\lambda^L$. This means that for a given energy scale the points inside the corresponding contour is allowed by this requirement. As can be seen from the figure, the position of the Landau poles is very similar to the situation for the $Z_2$-symmetric cases and there is only a small correlation between the values of the aligned Yukawas where the Landau pole is reached.

\begin{figure}
\hspace*{4cm}\includegraphics[width=6.2cm]{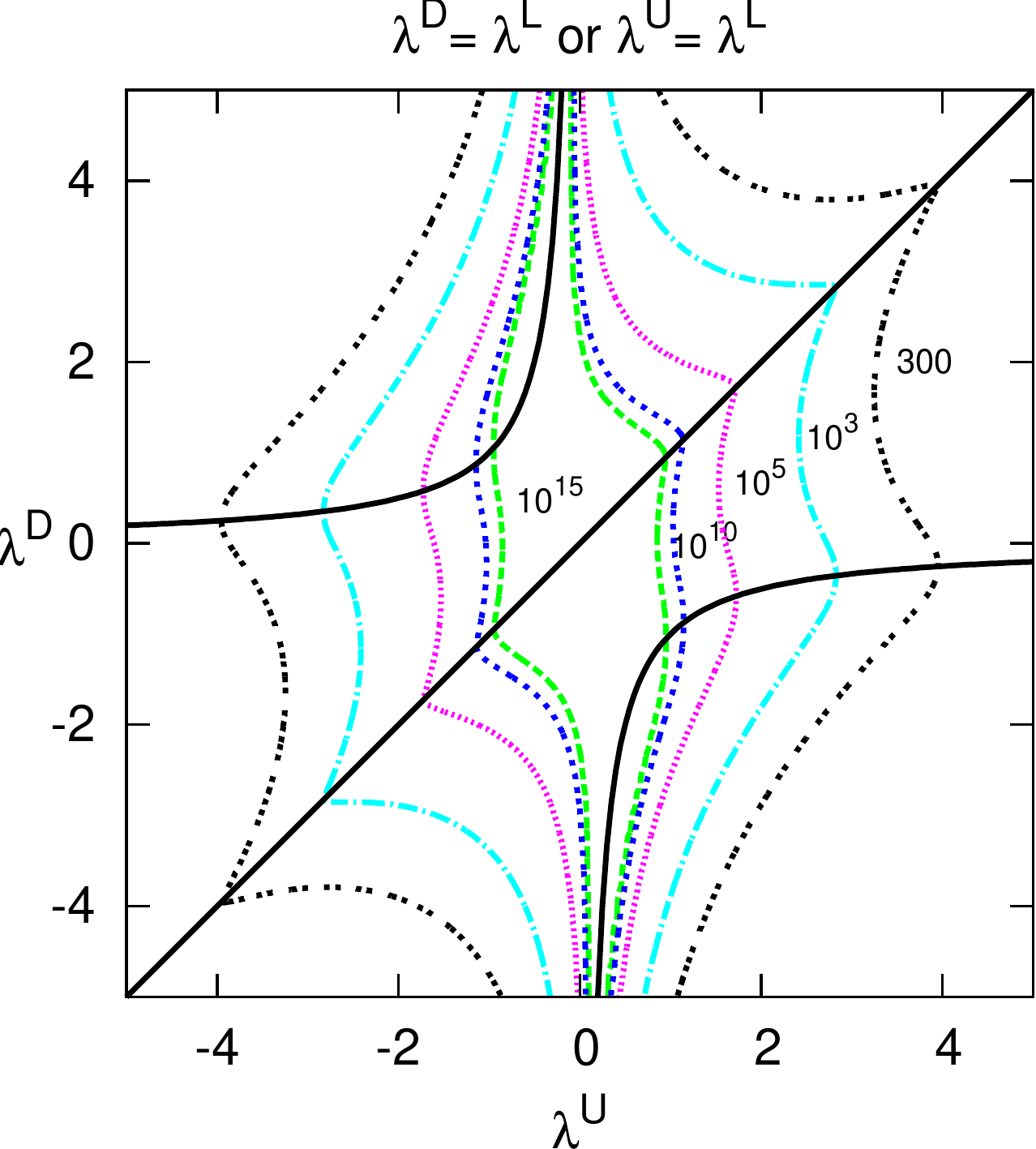}
  \caption{Same as Fig.~\ref{plot:2DA_landau} but also applying the constraints from the non-diagonal $\lambda^F_{i \neq j}$. The plot shows the results for $\lambda^L_{ii}=\lambda^D_{ii}$, but the same results are also obtained for $\lambda^L_{ii}=\lambda^U_{ii}$.}
\label{plot:2DA}
\end{figure}

Applying also the condition that the off-diagonal elements of should respect the limits given by the meson mixing constraints also at higher scales has a potentially large impact on the allowed regions. This is the case for the aligned models of type I/II and III, where $\lambda^L$ is set equal to $\lambda^D$ and $\lambda^U$ respectively, as can be seen in Fig.~\ref{plot:2DA}. In fact, within the parameter region  displayed in the figure  
(note the difference in scale compared to Fig.~\ref{plot:2DA_landau}) there is no difference between the two cases and therefore we only show one of them.
However, as may also have been expected, there are no additional constraints in the case when $\lambda^D$ and $\lambda^U$ are set equal since the off-diagonal lepton Yukawas are always small as a consequence of the small lepton masses and the limited cross-talk between the quarks and leptons.  In other words breaking the $Z_2$ symmetry between the quarks and leptons has no effect in this respect. 

\begin{figure}
\includegraphics[width=14cm, viewport=50 115 800 460,clip]{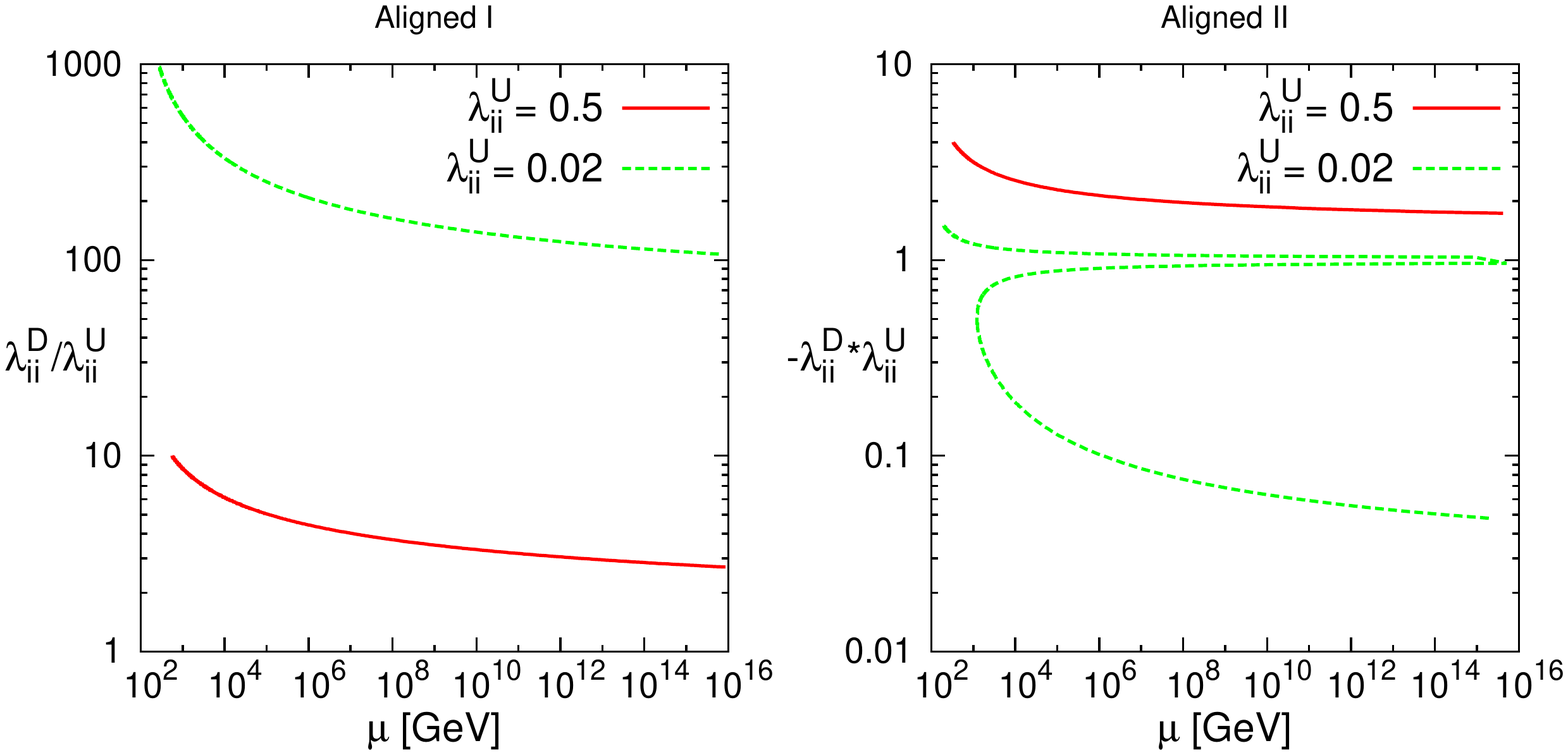}
  \caption{The constraints on the starting values of $\xi=\lambda^D_{ii}/\lambda^U_{ii}$ (left) and 
  $\xi=-\lambda^D_{ii} \lambda^U_{ii}$ (right)  as a function of the renormalization scale where the off-diagonal elements reaches 0.1 in the Aligned models of type I and type II respectively for the representative values $ \lambda^U_{ii}  =  0.02$ and  $0.5$.}
\label{plot:aligned}
\end{figure}

 For reference we have also included lines corresponding to the $Z_2$ symmetric relations in Fig.~\ref{plot:2DA}. Along these lines it is the Landau pole that gives the limit but in the other regions the limit comes from the off-diagonal elements. 
We also note that the plot is symmetric under  inversion through the origin $(x,y) \to (-x,-y)$, which follows since the evolution equations for $\rho^F_0$ are all even under $\rho^F_0 \to -\rho^F_0$ as long as the imaginary parts of $\kappa^F_0$ and $\rho^F_0$ are small.

It is also interesting to compare the results for non-equal  $\lambda^D$ and $\lambda^U$ with the constraints on $\lambda_{bb}$ and $\lambda_{tt}$ obtained from $b \to s \gamma$ in~\cite{Mahmoudi:2009zx}. Applying the conditions of stability when evolving to higher scales and that the non-diagonal Yukawas should stay small essentially removes the regions $|\lambda_{tt}| \gtrsim 1$ including the fine-tuned regions where  $\lambda_{bb}$ and $\lambda_{tt}$ are both large  ($\gtrsim 2$) and have the same sign.

As special cases we also show in Fig.~\ref{plot:aligned} the results for 
$ \lambda^U_{ii}  =  0.02, 0.5$ and either $\lambda^D_{ii}  = \lambda^L_{ii}  = \xi \lambda^U_{ii}$ (type I) or
$ \lambda^D_{ii}  = \lambda^L_{ii}  = -\xi/\lambda^U_{ii} $ (type II).
From these plots it is clear that for $\lambda^U_{ii}=0.5$, the 
off-diagonal elements puts strong constraints on the $Z_2$-symmetry breaking parameter
 $\xi=\lambda^D_{ii}/\lambda^U_{ii}$ ( $\xi=-\lambda^D_{ii}\lambda^U_{ii}$)
 with typical values being $\xi \lesssim 3-10 \, (2-5)$ for type I (II). For $\lambda^U_{ii}=0.02$ on the other hand the constraints are very mild in a type I set-up with $\xi \lesssim 100-1000$ allowed, whereas in a type II setup only $\xi$ values very close to 1 or  $\xi \lesssim  0.05 - 0.1 $ are allowed. The two possibilities corresponds to two distinct regions in the $\lambda^D_{ii} , \lambda^U_{ii}$ plane. The first one where $ \lambda^D_{ii} \approx -1/\lambda^U_{ii} $ and the second one where  $ \lambda^D_{ii}$ is small ($ \lesssim  2-5 $). For comparison we recall that the Landau pole constrains $\lambda^D_{ii} \lesssim 70-200$ more or less irrespectively of $\lambda^U_{ii} $. So the constraints on $\xi$ are more or less trivial in this case.

\subsubsection{Diagonal models}

Next we consider in more detail models with $Z_2$-breaking in either the up or the down sector. To make the discussion more clear we only consider models where $\lambda_{tt}$ and $\lambda_{bb}$ are related in a $Z_2$ symmetric way and since we have seen that the effects of the lepton sector is small we always set $\lambda^L_{ii}  = \lambda_{bb}$. (If  $\lambda_{tt}$ and $\lambda_{bb}$ are \emph{not} related in a $Z_2$ symmetric way then we are more or less back in the aligned models since these two are the dominant Yukawas). In other words we only partially break the  alignment.

\begin{figure}
\includegraphics[width=6.2cm]{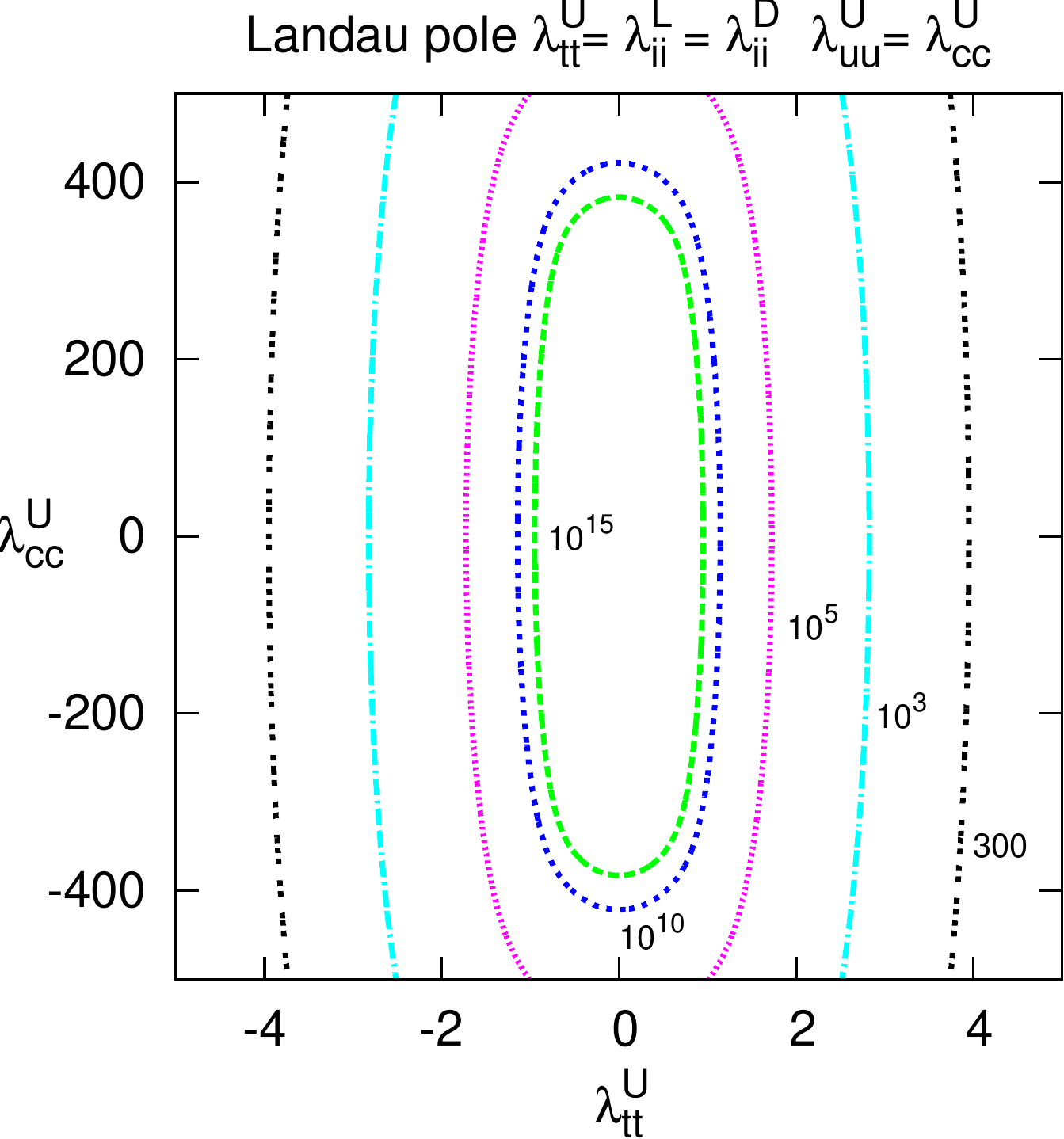} \hspace*{5mm}
\includegraphics[width=6.2cm]{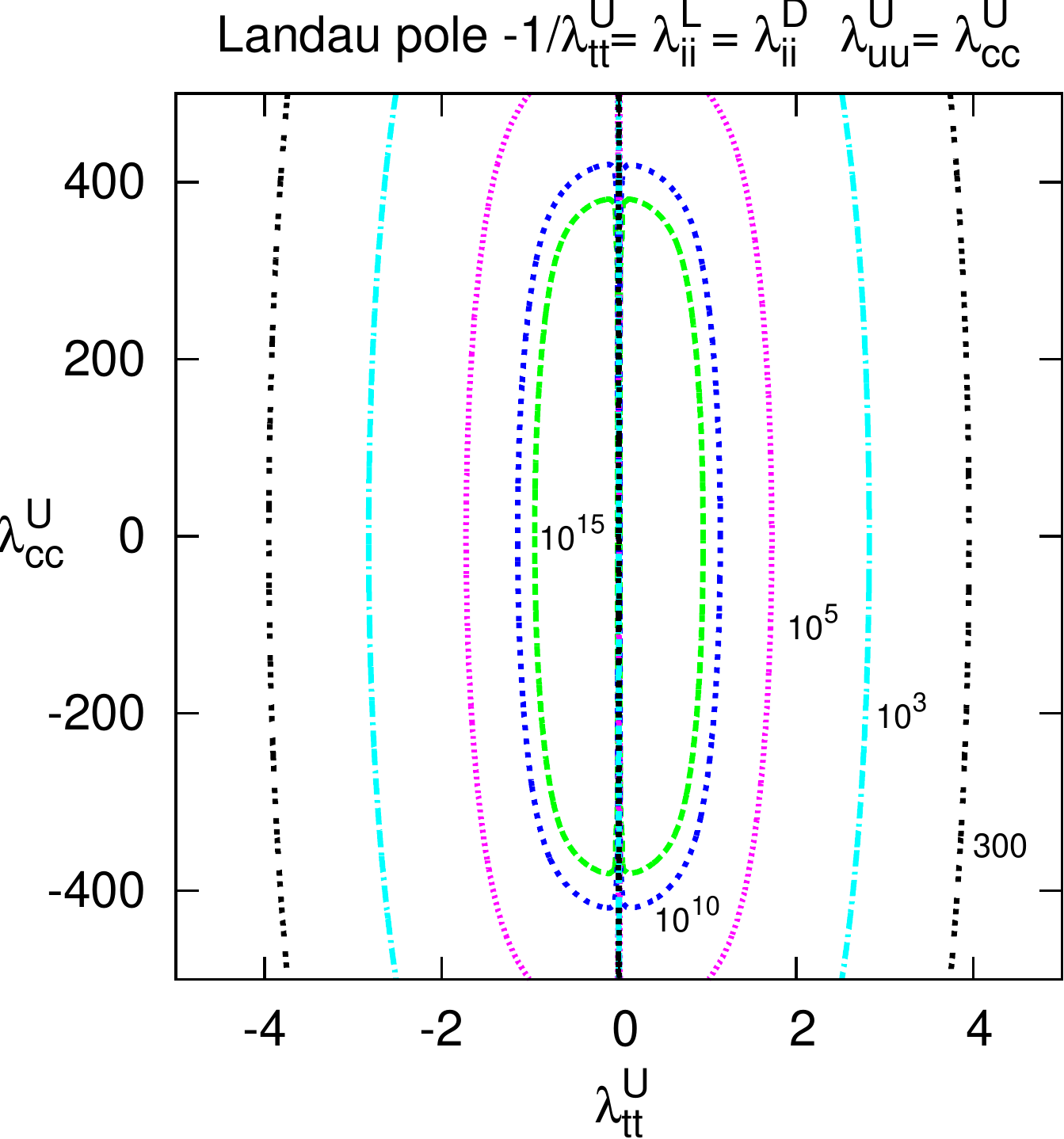} \newline \\
\includegraphics[width=6.2cm]{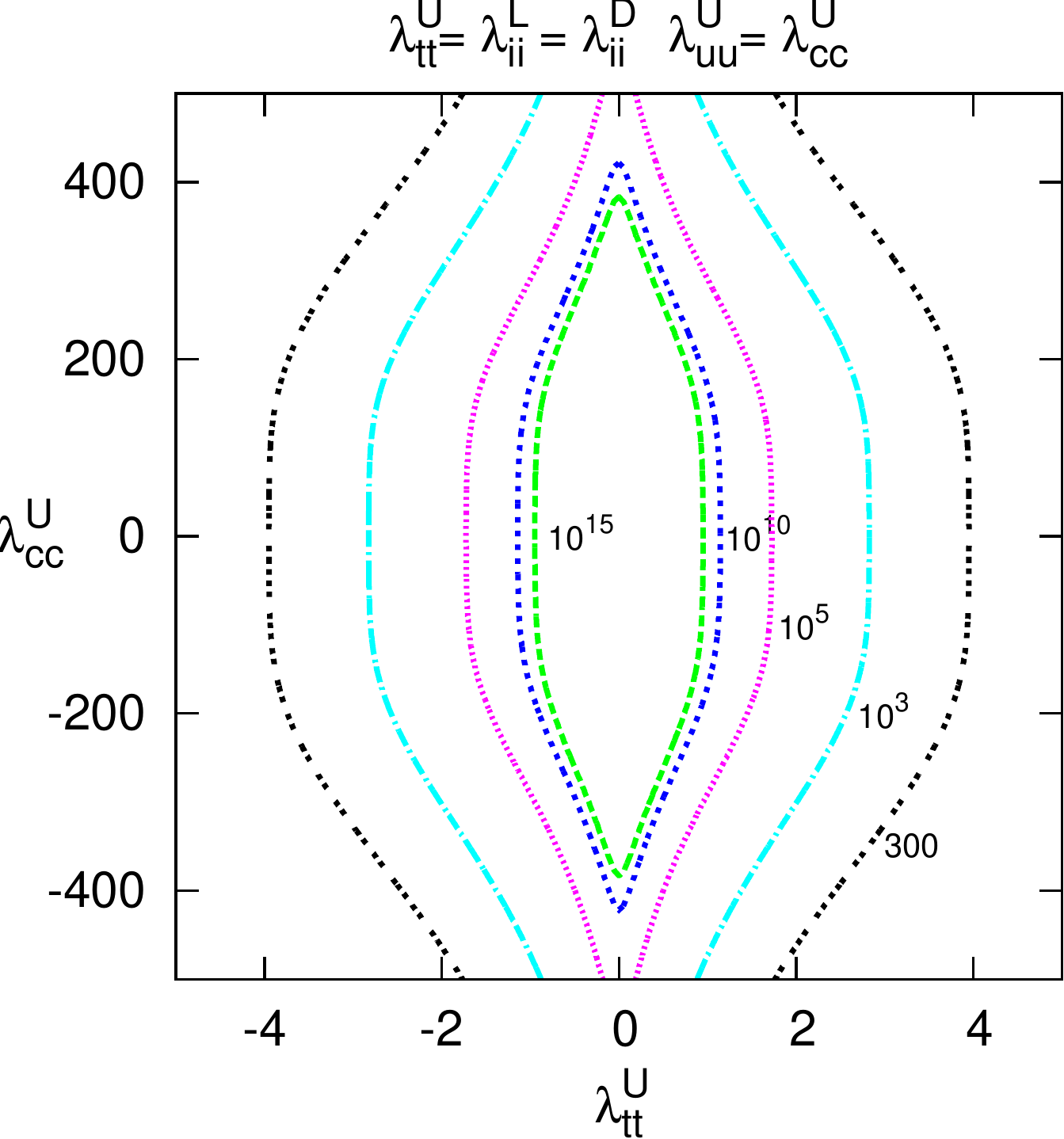} \hspace*{5mm}
\includegraphics[width=6.2cm]{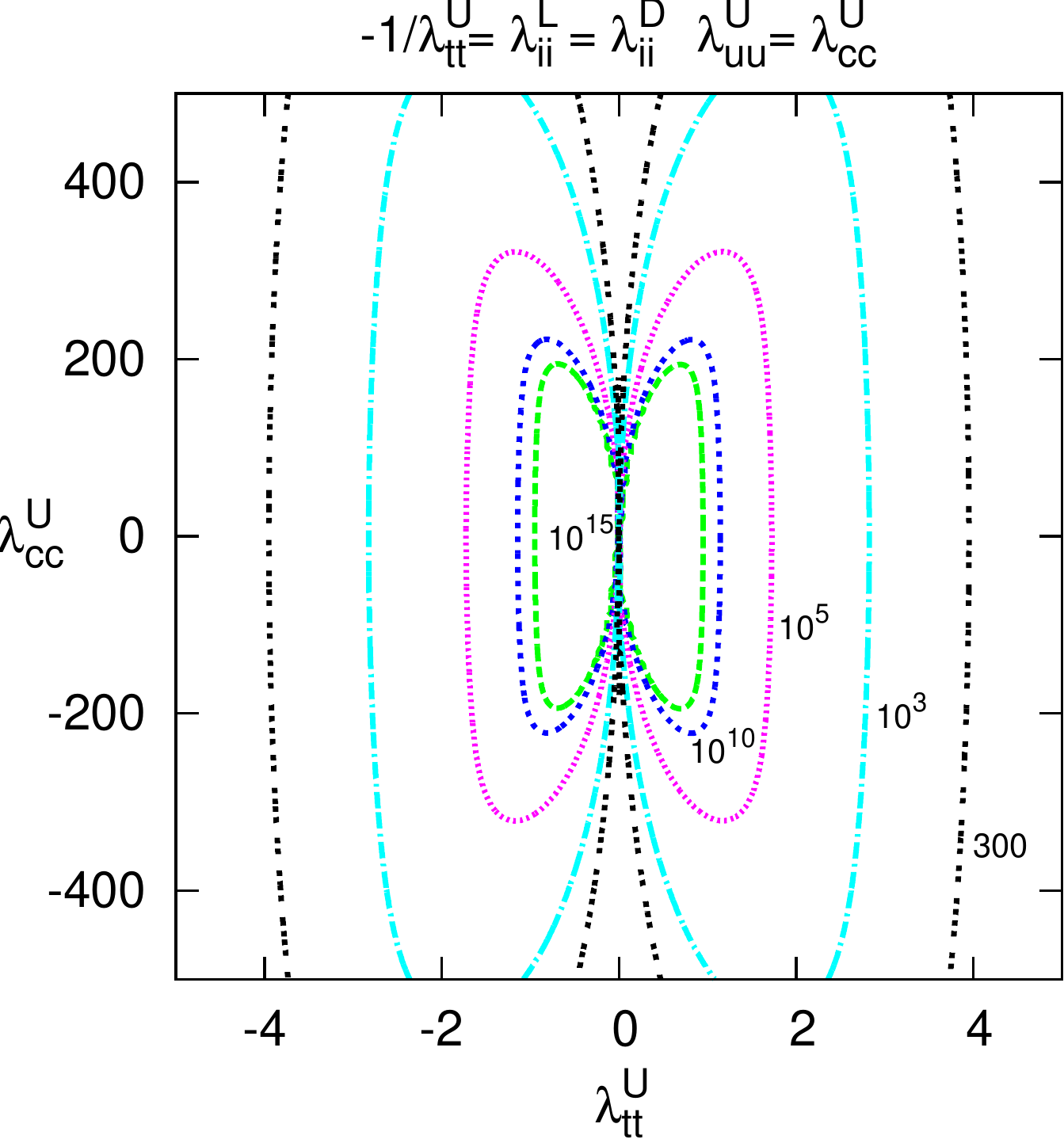}
  \caption{The energy scale where the Landau pole is reached (upper panels) together with the scale where one of the non-diagonal $ \lambda^F_{i \neq j}  =  0.1$ (lower panels) as a function of the input values $\lambda_{cc}$ and $\lambda_{tt}$. In the left (right) panels $\lambda^{D}=\lambda^{L}=\lambda_{tt} (-1/\lambda_{tt})$.}
\label{plot:2D_MSU}
\end{figure}

Thus we start with considering $Z_2$-breaking in the up-sector with  $\lambda^D=\lambda_{tt}$ (type I) or $\lambda^D=-1/\lambda_{tt}$ (type II). For simplicity we also set $ \lambda_{uu} =\lambda_{cc} $.

First of all, as we show in Fig.~\ref{plot:2D_MSU}, the Landau pole gives the restriction $  \lambda_{cc} \lesssim 400-500 $ both for type I and II, again more or less independently of the value of $\lambda_{tt}$. We also want to emphasize that even though it is not really discernable from the figure, there is also a lower limit on  $\lambda_{tt}\gtrsim 0.01$ from the Landau pole for $\lambda_{bb}$ for type II. 

The figure also shows that the impact of constraining the off-diagonal elements to be less than 0.1 is  limited for the type I set-up. In fact for $\lambda_{tt} = 0$ there is not additional constraint from the off-diagonal elements. In the type II set-up the constraints are more severe but even so quite mild. 

\begin{figure}
\includegraphics[width=14cm, viewport=50 115 800 460,clip]{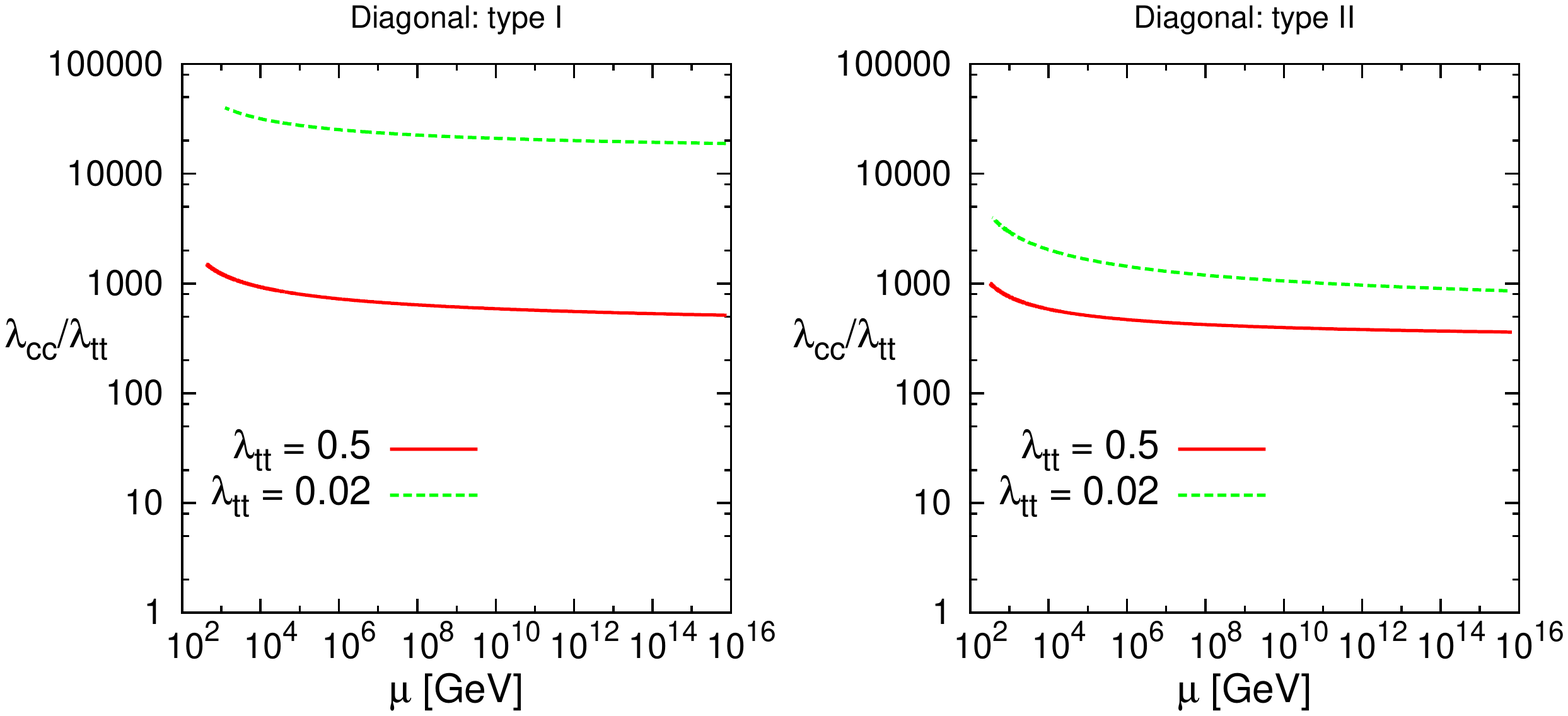}
  \caption{The constraints on the input values $\xi=\lambda_{cc}/\lambda_{tt}$  as a function of the renormalization scale where the off-diagonal elements reaches 0.1 in the diagonal models of type I (left) and type II (right) for the representative values $ \lambda_{tt}  =  0.02$ and  $0.5$.}
\label{plot:msu}
\end{figure}

To get a better picture of the range of the amount of $Z_2$-breaking allowed we  also give in Fig.~\ref{plot:msu} the constraints on the ratio $\lambda_{cc}/\lambda_{tt}$ in type I and II set ups for our standard values $ \lambda_{tt}  =  0.02$ and  $0.5$. From the plots it is clear that this ratio can be as large as $\sim1000$ without generating off-diagonal $\lambda^F \geq 0.1$ all the way up to the GUT scale. 

\begin{figure}
\includegraphics[width=6.2cm]{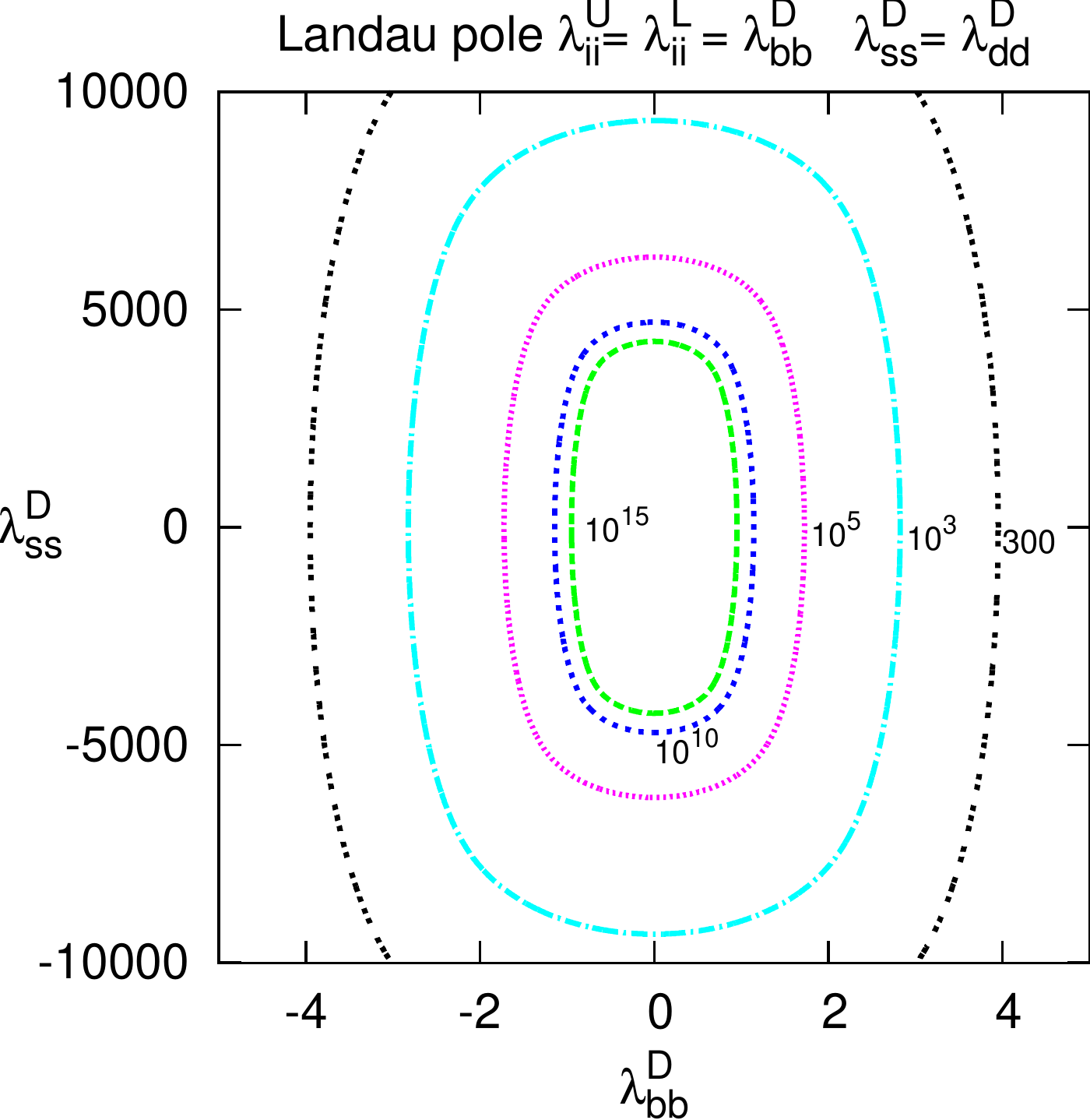} \hspace*{5mm}
\includegraphics[width=6.4cm]{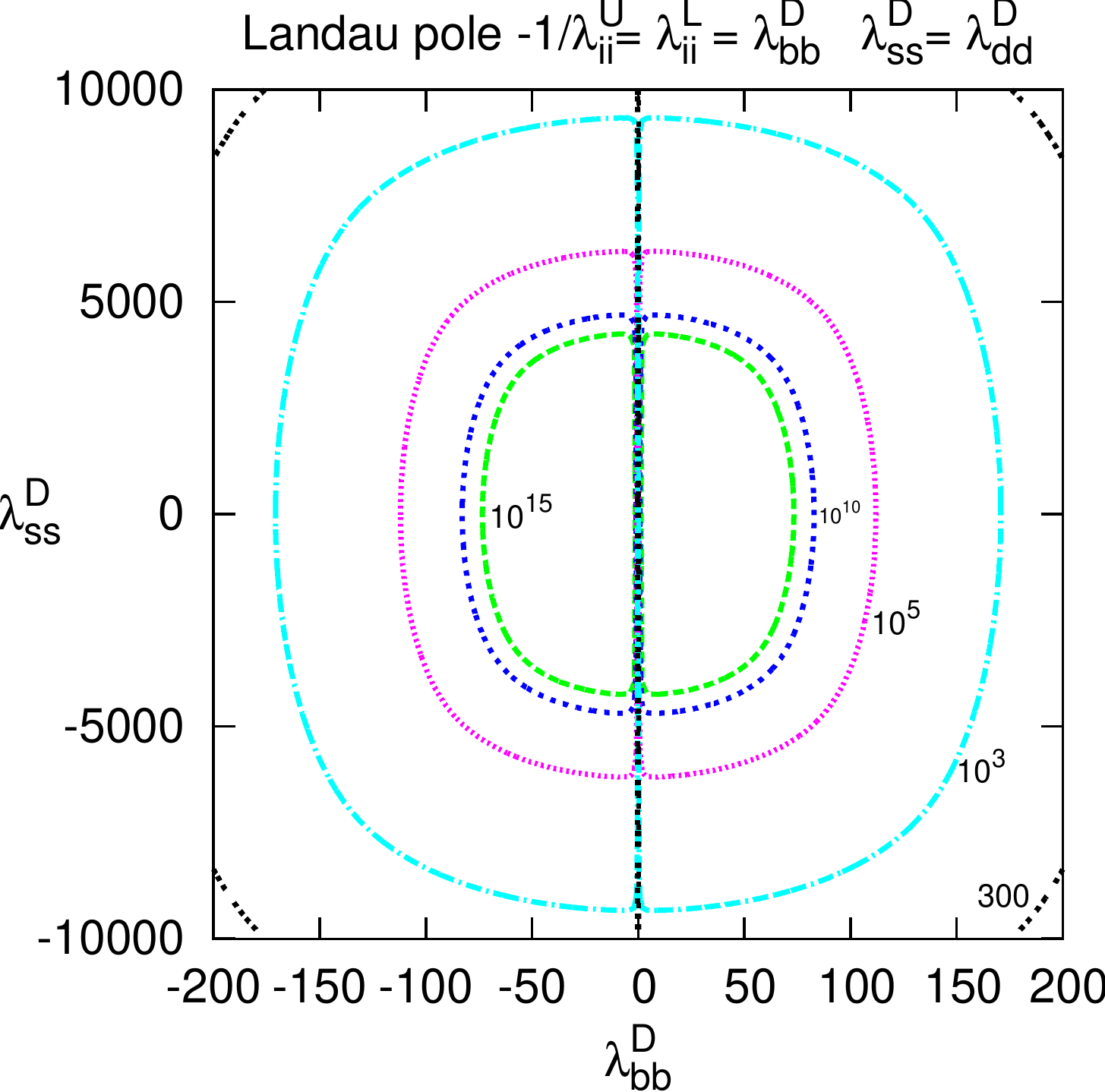} \newline \\
\includegraphics[width=6.2cm]{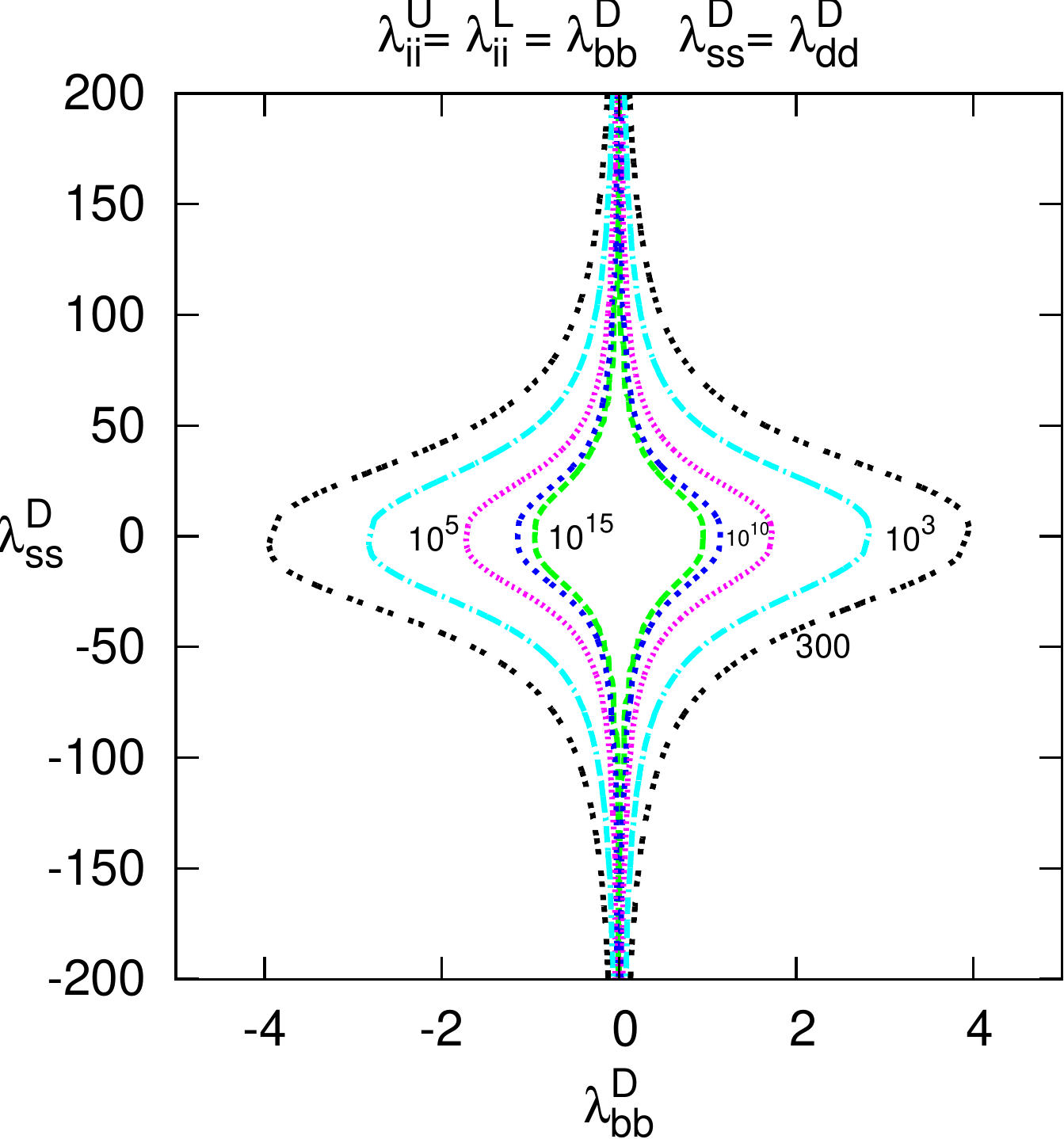} \hspace*{5mm}
\includegraphics[width=6.4cm]{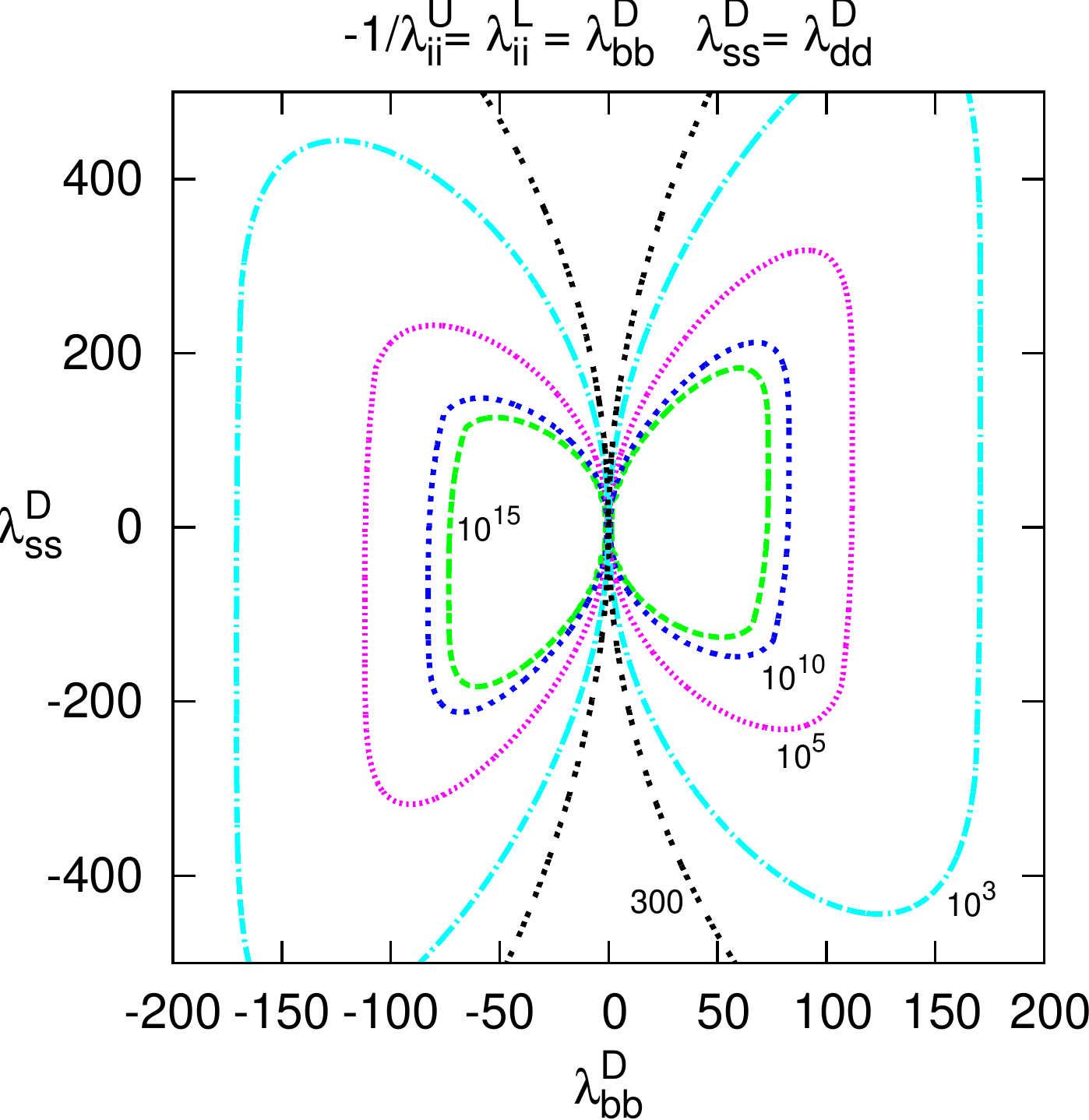}
  \caption{The energy scale where the Landau pole is reached (upper panels) together with the scale where one of the non-diagonal $ \lambda^F_{i \neq j}  =  0.1$ (lower panels) as a function of $\lambda_{ss}$ and $\lambda_{bb}$. In the left (right) panels $\lambda^{U}=\lambda_{bb} (-1/\lambda_{bb})$ and in all cases $\lambda^{L}=\lambda_{bb}$. }
\label{plot:2D_MSD}
\end{figure}

Next we consider $Z_2$-breaking in the down-sector with  $\lambda_{bb}=\lambda^U_{ii}$ (type I) or $\lambda_{bb}=-1/\lambda^U_{ii}$ (type II). Similarly to the up-sector we set $ \lambda_{dd} =\lambda_{ss} $ for simplicity. Also in this case the constraints from the Landau pole are similar for the two set-ups with $\lambda_{ss} \lesssim 400 -700$ in both cases with a small correlation with the value of $\lambda^U_{ii}$ and $\lambda_{bb}$ for a set up of type I and type II respectively as can be seen from Fig.~\ref{plot:2D_MSD} (upper panels). However, contrary to the up-sector the figure (lower panels) also shows that the effects from requiring the off-diagonal Yukawas to be small are quite severe. In the type II case one can even see a mild preference for solutions with  $ \lambda_{ss} \approx\lambda_{bb} $.

\begin{figure}
\includegraphics[width=14cm, viewport=50 115 800 460,clip]{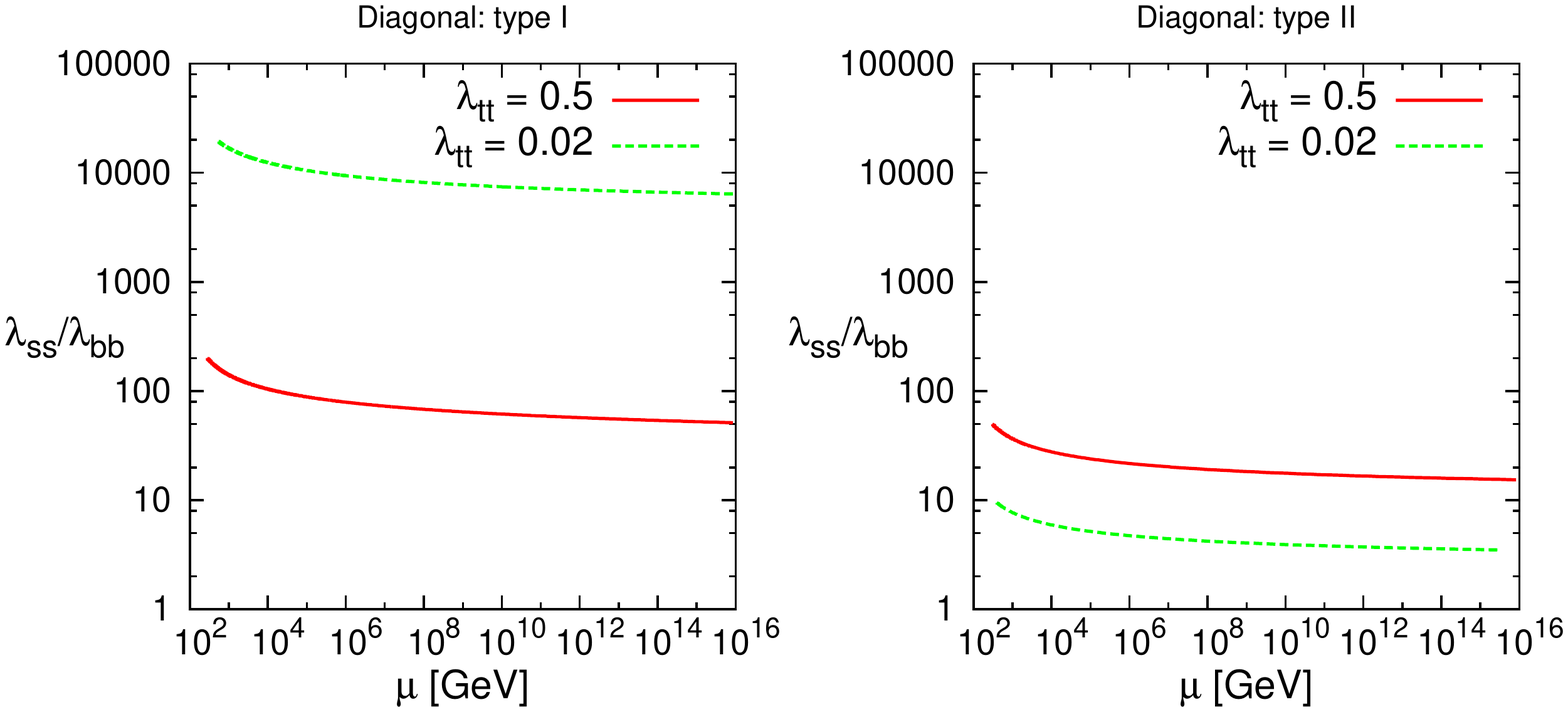}
  \caption{The constraints on the input values $\xi=\lambda_{ss}/\lambda_{bb}$  as a function of the renormalization scale where the off-diagonal elements reaches 0.1 in the diagonal models of type I (left) and type II (right) for the representative values $ \lambda_{tt}  =  0.02$ and  $0.5$.}
\label{plot:msd}
\end{figure}

To get a more quantitative picture of the constraints we show in Fig.~\ref{plot:msd} the ratio 
$\xi= \lambda_{ss} /\lambda_{bb} $ for type I and type II using the values $ \lambda^U_{ii}  =  0.02$ and  $0.5$. In the type II set-up the constraints are especially restrictive with $ \xi \lesssim 4-10 $ for  $ \lambda^U_{ii}  =  0.02$. In the type I set-up the constraints are less severe but even so stronger than the corresponding ones from the up-sector.

\subsubsection{Non-diagonal models}

\begin{figure}
\includegraphics[width=6.2cm]{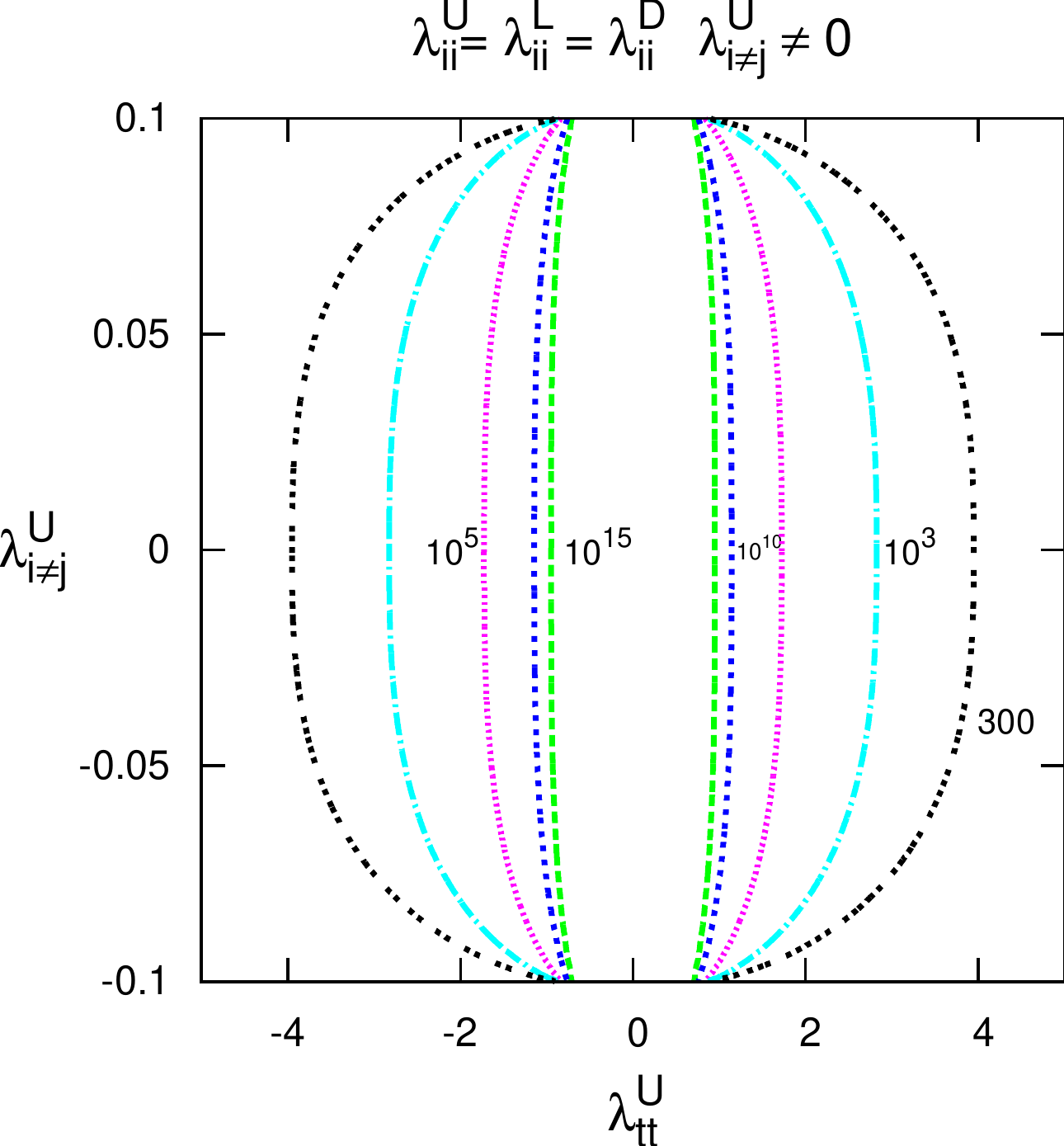} \hspace*{5mm}
\includegraphics[width=6.2cm]{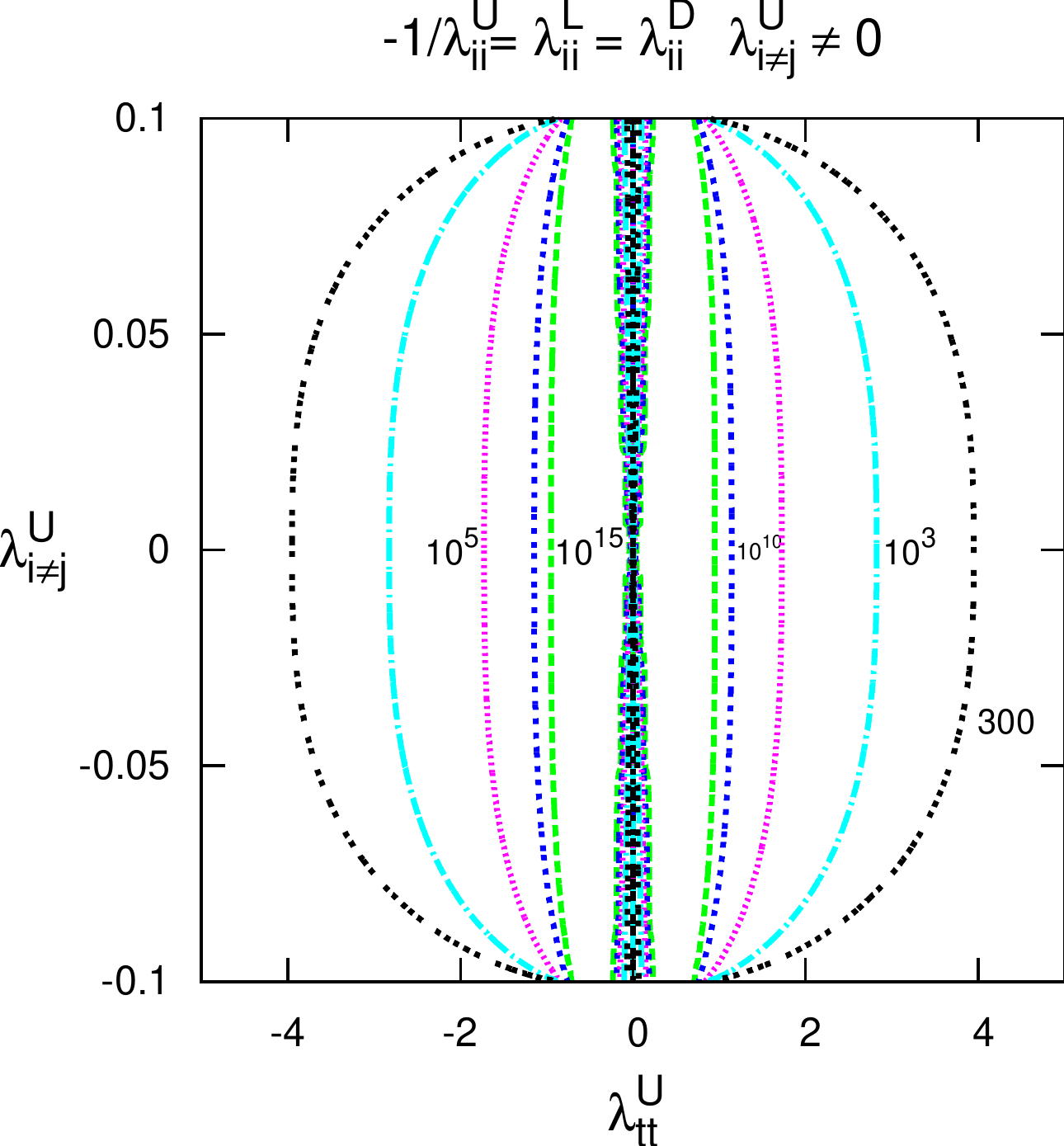} \newline \\
\includegraphics[width=6.2cm]{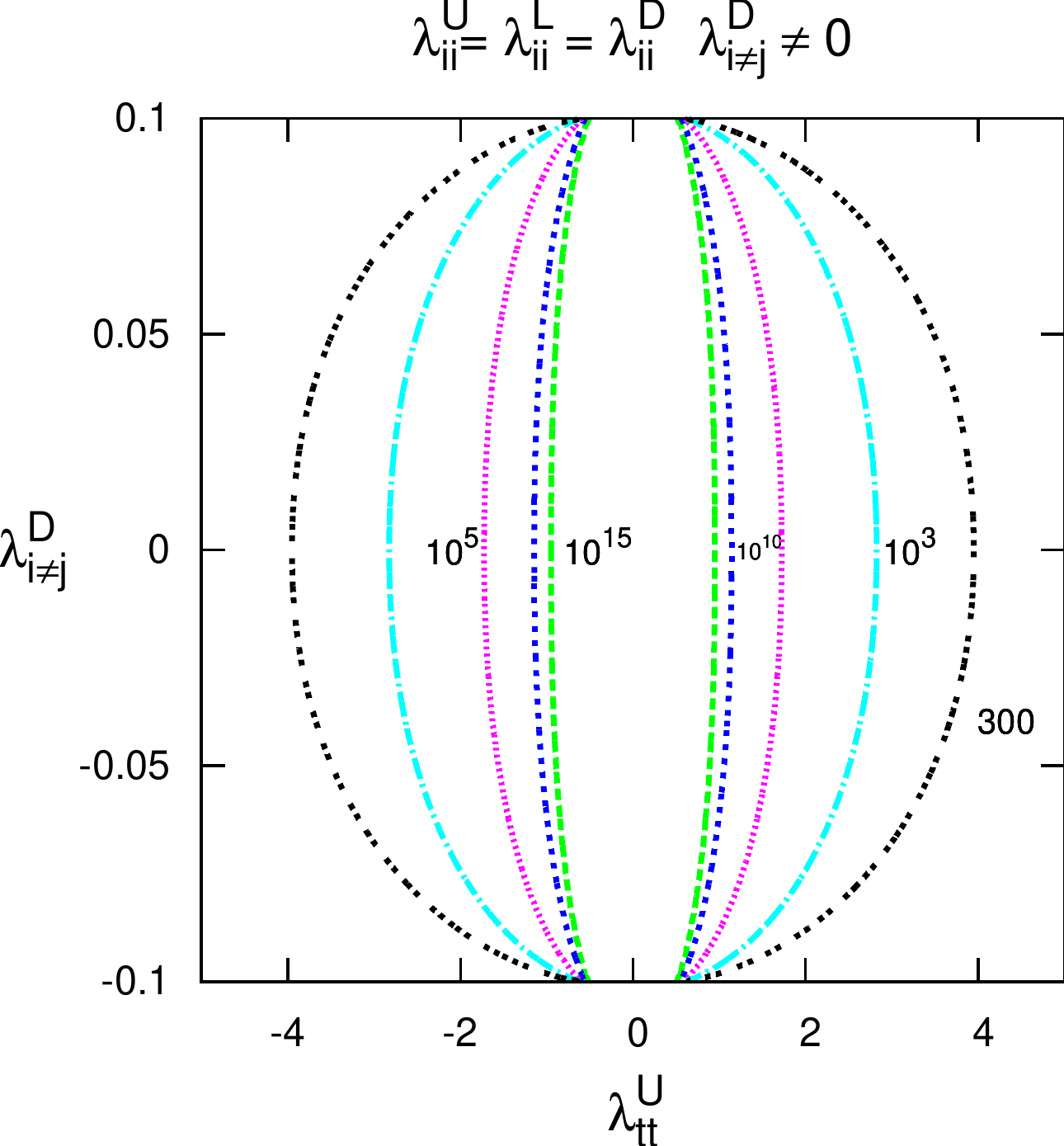} \hspace*{5mm}
\includegraphics[width=6.2cm]{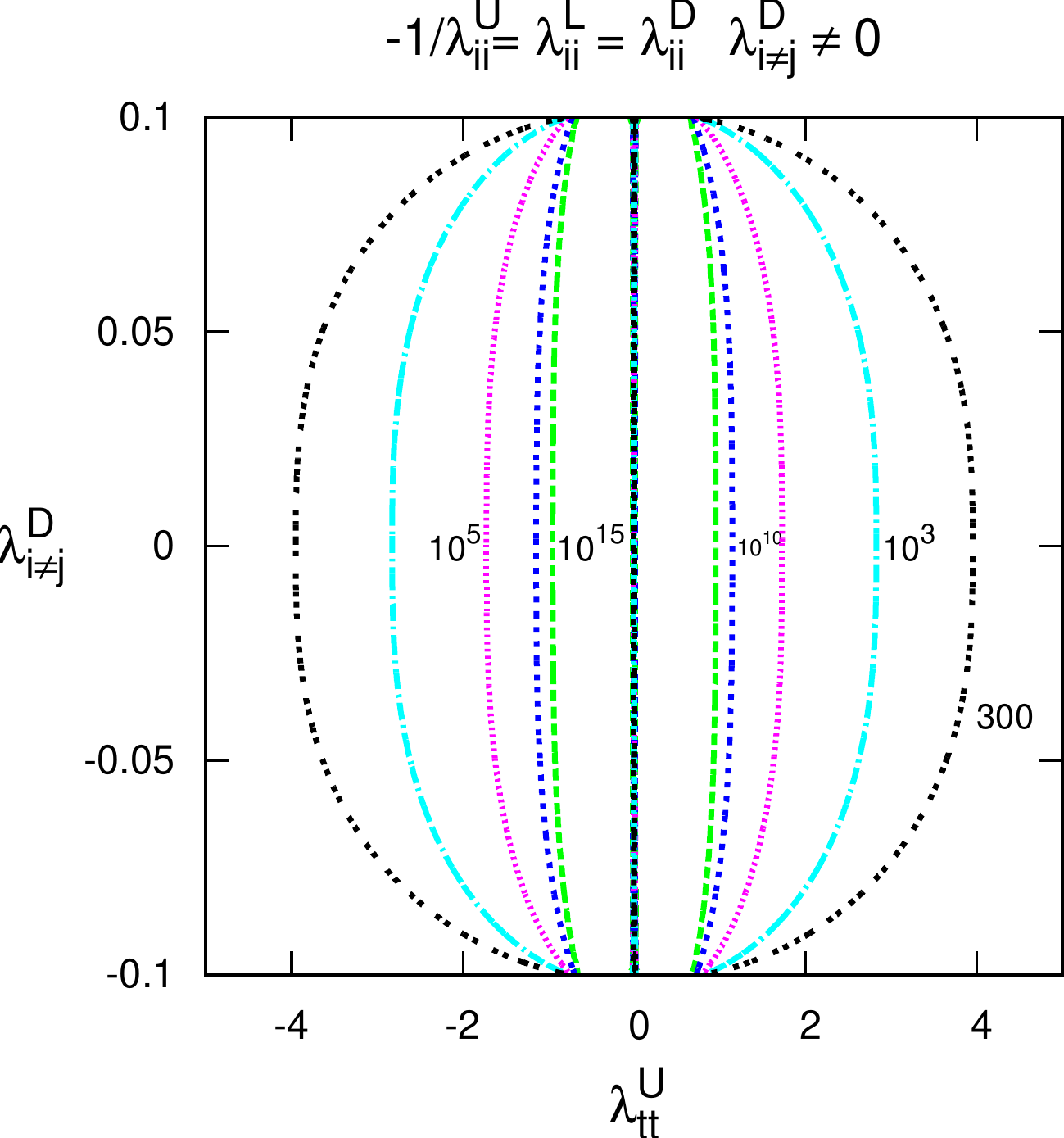}
  \caption{The constraints from the Landau pole and the off-diagonal elements as a function of $\lambda^U_{ii}$ and the off-diagonal elements $\lambda^U_{i \neq j}$ (up) or    $\lambda^D_{i \neq j}$ (down)  at the input scale for the type I (left) and type II (right) relations for the diagonal elements.}
\end{figure}

Finally we consider the case of breaking the $Z_2$-symmetry from having non-zero non-diagonal elements in the up- or down sectors.
As starting point we again use the $Z_2$ symmetric models of type I or II for the diagonal elements and then set either  
$\lambda^U_{i \neq j}=0.1 $ or $\lambda^D_{i \neq j}=0.1$  at the EW scale in order to break the  $Z_2$ symmetry.

\begin{figure}
        \includegraphics[width=7.5cm, viewport=140 20 700 550,clip]{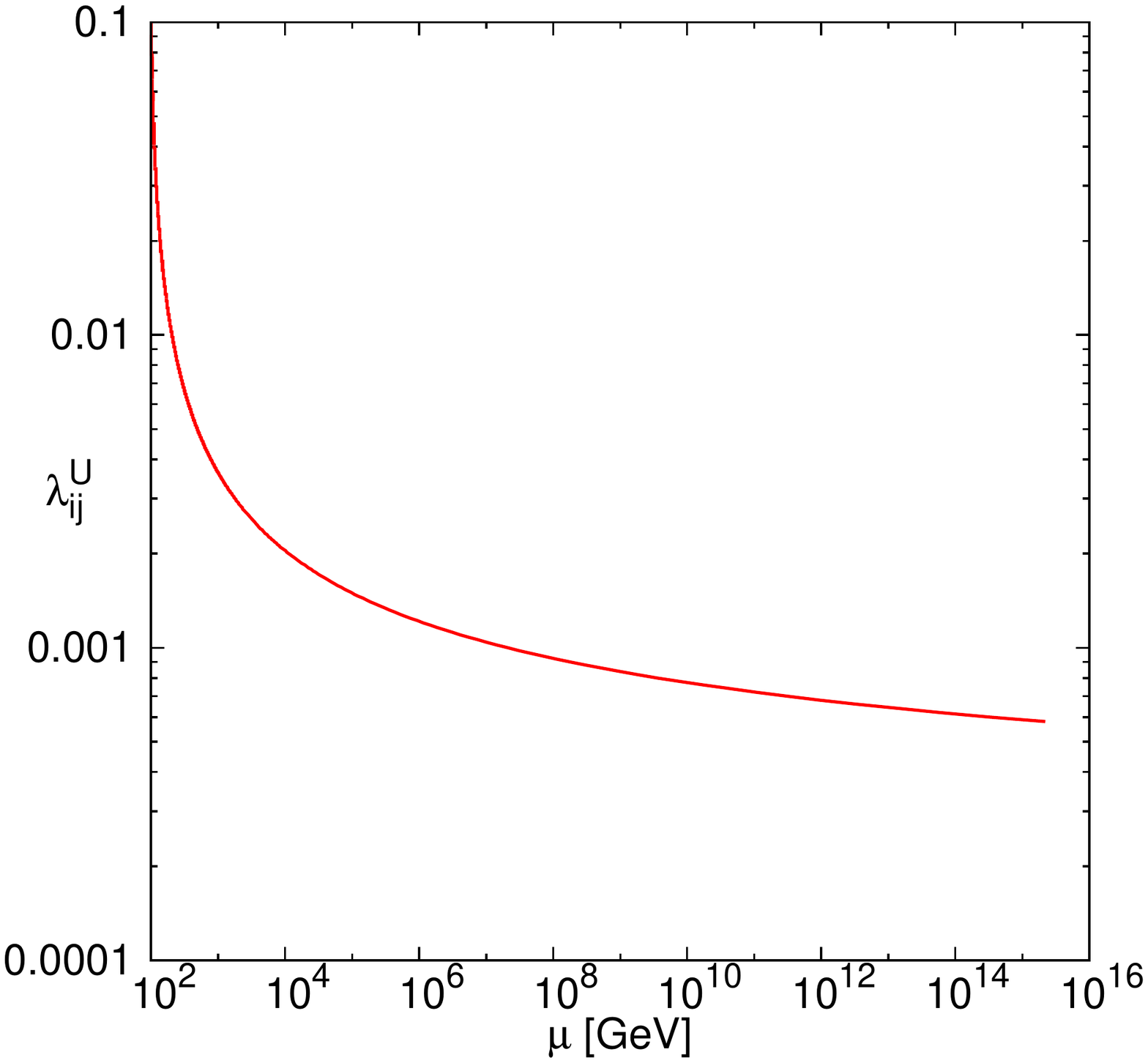}
 \includegraphics[width=7.5cm]{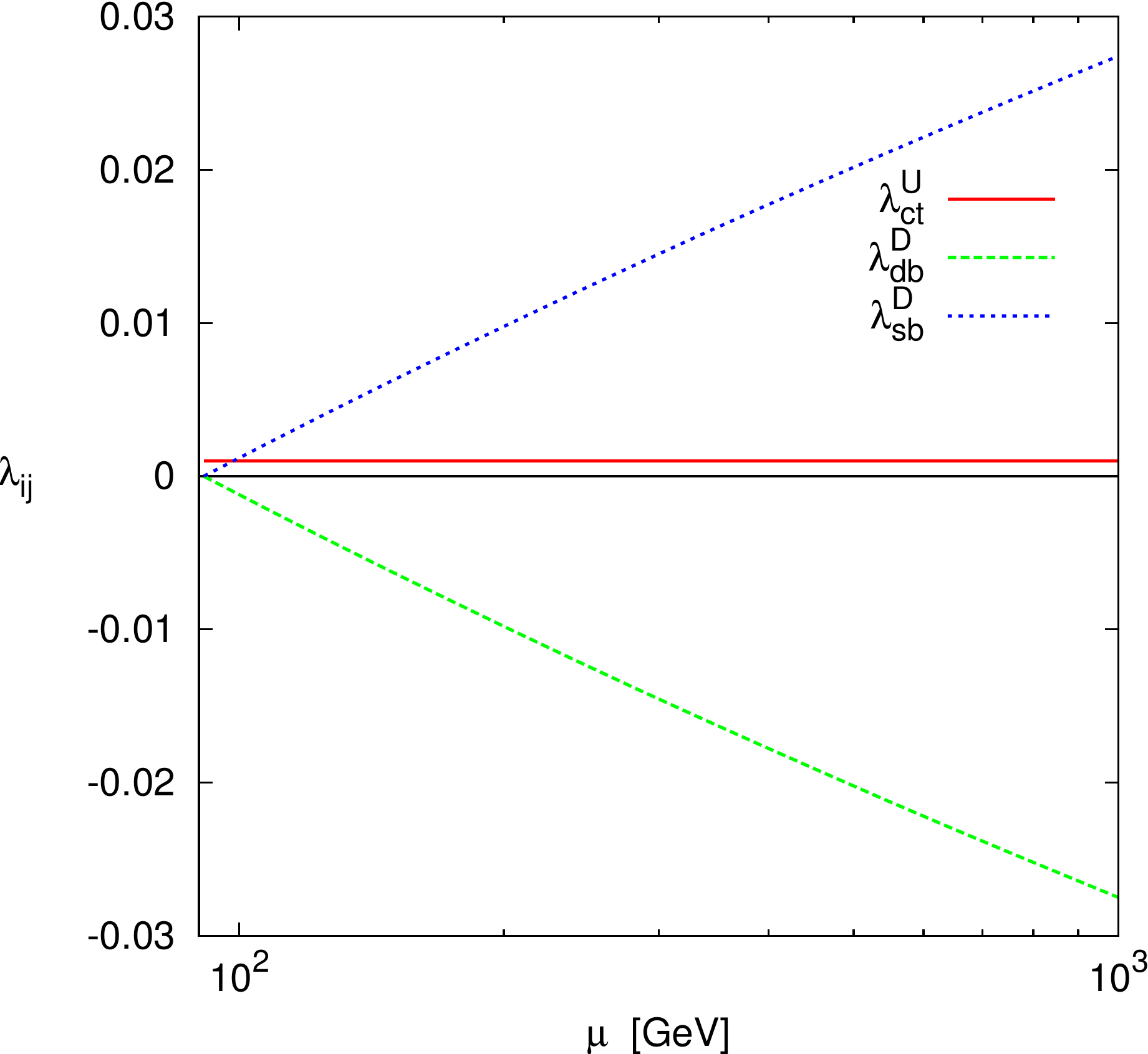}
 \caption{Left: The starting value $\lambda^U_{i \neq j}$ as a function of the energy scale $\mu$ where one of the non-diagonal elements of $\lambda^{U,D}_{i \neq j}$ becomes larger than $0.1$ for $\lambda^U_{ii} = 0.02$ and a type II relation between the diagonal elements, $ \lambda^D_{ii}  = \lambda^L_{ii}  = -1/\lambda^U_{ii} $. Right: The RGE-evolution of  the non-diagonal elements $\lambda_{ct}(\mu)$,  $\lambda_{sb}(\mu)$, and $\lambda_{db}(\mu)$   in the same case for $\lambda^U_{i \neq j}=0.001$. }
\label{plot:NDcs}
\end{figure}

Quite unexpectedly the additional constraints from requiring the off-diagonal elements to stay small are limited. The corresponding plots for the case of only considering the Landau pole are essentially straight vertical lines. Thus we do not show the effects of applying the two constraints separately. 
In fact it is only in case II with $ \lambda^D_{ii}  = \lambda^L_{ii}  = -1/\lambda^U_{ii} $ and $\lambda^U_{i \neq j}=0.1 $ that the requirement of having $\lambda^U_{i \neq j}(\mu)\leq0.1 $ gives any discernable effect and then only for small $\lambda^U_{ii} \lesssim 0.2$. On the other hand, in this case the constraints are very strong as also illustrated in Fig.~\ref{plot:NDcs}. It is interesting to note that it is actually the off-diagonal elements in the down-sector that become large whereas the ones in the up-sector remain in accord with the limit $\lambda^U_{i \neq j}(\mu)\leq0.1 $. This means that even though there are presently no direct experimental constraints on $\lambda_{ct}$ and $\lambda_{ut}$ they are in this case highly constrained from the link to the down-sector through the RGE evolution. This is then the case in the MSSM, the prime example of a type II 2HDM, for large $\tan\beta$.
To see more clearly what happens we show also in Fig.~\ref{plot:NDcs} the RGE evolution of the relevant off-diagonal elements for the input values  $\lambda^U_{ii} = 0.02$ and $\lambda^U_{i \neq j}=0.001 $, $\lambda^D_{i \neq j}=0 $.

\section{Conclusion}
\label{sec:conclusion}

We have seen that the RGE evolution is a useful tool to analyze the stability of the assumptions underlying different versions of the 2HDM under variations of the scale where the model is defined. A large sensitivity indicates 
that the assumptions behind the model are not stable meaning that they are either fine-tuned or incomplete such that there for example will be additional particles appearing when going to a higher energy. From this respect we have studied both the appearance of a Landau pole as well as off-diagonal Yukawa couplings leading to FCNC larger than experimentally allowed at the EW scale.

Based on our studies we have seen that the constraints from avoiding a Landau-pole are in general the same irrespective of the $Z_2$-symmetry. 
They appear as soon as the magnitude of one of the Yukawa couplings becomes of order 1.

The constraints from the off-diagonal elements on the other hand depend on the details of how the $Z_2$-symmetry is broken:
\begin{itemize}
\item
breaking the $Z_2$ relation between $\lambda^D$ and $\lambda^U$ as in the Aligned models is highly constrained with 
$\lambda^D/\lambda^U \lesssim 10$ or $-\lambda^D\lambda^U \lesssim 10$ 
unless  $\lambda^D$ and $\lambda^U$ are both $\lesssim 2$,
\item
breaking it instead in the up-sector by having  $\lambda_{cc}$ and $\lambda_{tt}$ non-equal gives a small difference compared to the constraints coming from the Landau pole with ratios  $\lambda_{cc}/\lambda_{tt} > 1000$ allowed,
\item
in the down sector the constraints can be much stronger, but also more dependent on the relation between $\lambda_{bb}$ and $\lambda_{tt}$, ranging from 
$\lambda_{ss} / \lambda_{bb} \lesssim 10$ for $\lambda_{bb}=50$ and $\lambda_{tt}=-0.02$ 
to 
$\lambda_{ss} / \lambda_{bb} \lesssim 10000$ for $\lambda_{bb}=\lambda_{tt}=0.02$,
\item
in the non-diagonal case the constraints are weak when starting from $\lambda^{D,U}_{i \neq j} = 0.1 $ except in the case $\lambda_{bb}=-50$ and $\lambda_{tt}=0.02$ where we find  $\lambda^U_{i \neq j} \lesssim 0.001 $. In all cases it is the  $\lambda^D_{i \neq j}$  that become large.
\end{itemize}

From this we can conclude that starting from a type I $Z_2$ symmetry there is quite a lot of room for breaking the symmetry as long as one does not encounter a Landau pole except that $\lambda^D/\lambda^U \lesssim 10$ has to be respected. In the type II case however, the room for breaking the symmetry is much smaller for large $\lambda_{bb}$. This is natural since in the latter case $\kappa_{tt}$ and $\rho_{bb}$ are both large. It is also interesting to note that this corresponds to the situation in the MSSM with large $\tan\beta$. Finally we conclude that there is little hope to see effects of non-diagonal Yukawa couplings in the top system in a type II model such as MSSM if $\tan\beta$ is large.

\section*{Acknowledgements}
This work is supported in part by the European Community-Research Infrastructure
Integrating Activity “Study of Strongly Interacting Matter” (HadronPhysics2, Grant
Agreement n. 227431) and the Swedish Research Council grants 621-2008-4074, 621-2008-4219 and
621-2010-3326.



\begin{thebibliography}{99}

\bibitem{Lee:1973iz}
 T.~D.~Lee,
 Phys.\ Rev.\  D {\bf 8} (1973) 1226.

\bibitem{Glashow:1976nt}
  S.~L.~Glashow, S.~Weinberg,
  Phys.\ Rev.\  {\bf D15 } (1977)  1958.

\bibitem{Pich:2009sp}
 A.~Pich and P.~Tuzon,
 Phys.\ Rev.\  D {\bf 80} (2009) 091702
 [arXiv:0908.1554].

\bibitem{Ferreira:2010xe}
  P.~M.~Ferreira, L.~Lavoura and J.~P.~Silva,
  Phys.\ Lett.\  B {\bf 688} (2010) 341
  [arXiv:1001.2561].

\bibitem{arXiv:1005.5310}
  A.~J.~Buras, M.~V.~Carlucci, S.~Gori and G.~Isidori,
  JHEP\ {\bf 1010} (2010) 009
  [arXiv:1005.5310].

\bibitem{arXiv:1005.5706}
  C.~B.~Braeuninger, A.~Ibarra and C.~Simonetto,
  Phys.\ Lett.\ B\ {\bf 692} (2010) 189
  [arXiv:1005.5706].


\bibitem{arXiv:1006.0470}
  M.~Jung, A.~Pich and P.~Tuzon,
  JHEP\ {\bf 1011} (2010) 003
  [arXiv:1006.0470].

\bibitem{Mahmoudi:2009zx}
  F.~Mahmoudi, O.~Stal,
  Phys.\ Rev.\  {\bf D81 } (2010)  035016.
  [arXiv:0907.1791].

\bibitem{Cheng:1987rs}
 T.~P.~Cheng and M.~Sher,
 Phys.\ Rev.\  D {\bf 35} (1987) 3484.


 \bibitem{cvetic:1997}
  G.~Cvetic, S.~S.~Hwang and C.~S.~Kim,
  Int.\ J.\ Mod.\ Phys.\  A {\bf 14}, 769 (1999)
  [arXiv:hep-ph/9706323].

\bibitem{Branco:2011iw}
  G.~C.~Branco, P.~M.~Ferreira, L.~Lavoura, M.~N.~Rebelo, M.~Sher, J.~P.~Silva,
  [arXiv:1106.0034].

\bibitem{hep-ph/0503172}
  A.~Djouadi,
  Phys.\ Rept.\ \ {\bf 457} (2008) 1
  [hep-ph/0503172].

\bibitem{hep-ph/0503173}
  A.~Djouadi,
  Phys.\ Rept.\ \ {\bf 459} (2008) 1
  [hep-ph/0503173].

\bibitem{hep-ph/0504050}
  S.~Davidson and H.~E.~Haber,
  Phys.\ Rev.\ D\ {\bf 72} (2005) 035004
   [Erratum-ibid.\ D\ {\bf 72} (2005) 099902]
  [hep-ph/0504050].
  
\bibitem{Gupta:2009wn}
  R.~S.~Gupta, J.~D.~Wells,
  Phys.\ Rev.\  {\bf D81 } (2010)  055012.
  [arXiv:0912.0267].

\bibitem{Ferreira:2009jb}
  P.~M.~Ferreira and D.~R.~T.~Jones,
  JHEP {\bf 0908} (2009) 069
  [arXiv:0903.2856 [hep-ph]].
  
\bibitem{Eriksson:2009ws}
  D.~Eriksson, J.~Rathsman and O.~Stal,
  Comput.\ Phys.\ Commun.\  {\bf 181} (2010) 189
  [arXiv:0902.0851 [hep-ph]].

\bibitem{hep-ph/0602242}
  H.~E.~Haber and D.~O'Neil,
  Phys.\ Rev.\ D\ {\bf 74} (2006) 015018
  [hep-ph/0602242].

\bibitem{Atwood:1996vj}
  D.~Atwood, L.~Reina, A.~Soni,
  Phys.\ Rev.\  {\bf D55 } (1997)  3156-3176.
  [hep-ph/9609279].

\bibitem{pdg}
  K.~Nakamura {\it et al.} [ Particle Data Group Collaboration ],
  J.\ Phys.\ G {\bf G37}, 075021 (2010).

\bibitem{Laiho:2009eu}
  J.~Laiho, E.~Lunghi, R.~S.~Van de Water,
  Phys.\ Rev.\  {\bf D81 } (2010)  034503.
  [arXiv:0910.2928].

\bibitem{Lunghi:2007ak}
  E.~Lunghi, A.~Soni,
  JHEP {\bf 0709 } (2007)  053.
  [arXiv:0707.0212].

\bibitem{Lenz:2010gu}
  A.~Lenz, U.~Nierste, J.~Charles, S.~Descotes-Genon, A.~Jantsch, C.~Kaufhold, H.~Lacker, S.~Monteil {\it et al.},
  Phys.\ Rev.\  {\bf D83 } (2011)  036004.
  [arXiv:1008.1593].

\bibitem{LHCb}
 LHCb collaboration, ``Measurement of $\Delta m_s$ in the decay $B^0_s \to D_s^- (K^+K^-\pi^-)\pi^+$ using opposite-side and same-side flavour tagging algorithms,''  LHCb-CONF-2011-050


\bibitem{Xing:2007fb}
  Z.~z.~Xing, H.~Zhang and S.~Zhou,
  Phys.\ Rev.\  D {\bf 77} (2008) 113016
  [arXiv:0712.1419].

\bibitem{eigen}
G.\ Guennebaud, B.\ Jacob  {\it et al.}, \url{http://eigen.tuxfamily.org/}


\bibitem{gsl}
M. Galassi, J. Davies, J. Theiler, B. Gough, G. Jungman, P. Alken, M. Booth, F. Rossi
"GNU Scientific Library Reference Manual - Third Edition (v1.12)",
\url{http://www.gnu.org/s/gsl/}

\end{thebibliography}
\end{document}